\documentclass[%
 reprint,
 superscriptaddress,
 amsmath,amssymb,
 aps,
]{revtex4-1}

\usepackage{graphicx}
\usepackage{bm}
\usepackage{braket}
\usepackage{comment}
\usepackage{amsthm}
\usepackage{multirow}
\usepackage{hyperref}

\newcommand{\average}[1]{\ensuremath{\langle#1\rangle} }
\def\vector#1{\mbox{\boldmath $#1$}}

\begin{document}

\preprint{APS/123-QED}

\title{Ferroelectricity-induced multiorbital odd-frequency superconductivity in SrTiO$_3$}

\author{Shota Kanasugi}
\email{kanasugi.shouta.62w@st.kyoto-u.ac.jp}
 \affiliation{%
 Department of Physics, Kyoto University, Kyoto 606-8502, Japan
}%
\author{Dushko Kuzmanovski}
\affiliation{%
 Nordita, KTH Royal Institute of Technology and Stockholm University, Roslagstullsbacken 23, SE-106 91 Stockholm, Sweden
}%
\author{Alexander V. Balatsky}
\affiliation{%
 Nordita, KTH Royal Institute of Technology and Stockholm University, Roslagstullsbacken 23, SE-106 91 Stockholm, Sweden
}%
\affiliation{%
 Department of Physics, University of Connecticut, Storrs, Connecticut 06269, USA
}%
\author{Youichi Yanase}%
\affiliation{%
 Department of Physics, Kyoto University, Kyoto 606-8502, Japan
}%
\affiliation{%
 Institute for Molecular Science, Okazaki 444-8585, Japan
}%

\date{\today}

\begin{abstract}
We demonstrate that SrTiO$_3$ can be a platform for observing the bulk odd-frequency superconducting state owing to the multiorbital/multiband nature. 
We consider a three-orbital tight-binding model for SrTiO$_3$ in the vicinity of a ferroelectric critical point. 
Assuming an intraorbital spin-singlet $s$-wave superconducting order parameter, it is shown that the odd-frequency pair correlations are generated due to the intrinsic LS coupling which leads to the local orbital mixing. 
Furthermore, we show the existence of additional odd-frequency pair correlations in the ferroelectric phase, which is induced by an odd-parity orbital hybridization term proportional to the ferroelectric order parameter. 
We also perform a group theoretical classification of the odd-frequency pair amplitudes based on the fermionic and space group symmetries of the system. 
The classification table enables us to predict dominant components of the odd-frequency pair correlations based on the symmetry of the normal state Hamiltonian that we take into account. 
Furthermore, we show that experimental signatures of the odd-parity orbital hybridization, which is an essential ingredient for the ferroelectricity-induced odd-frequency pair correlations, can be observed in the spectral functions and density of states. 
\end{abstract}

\pacs{Valid PACS appear here}
\maketitle


\section{\label{sec1} Introduction}
The Berezinskii rule \cite{AVB92}  is a statement about the permutation anti-symmetry of the anomalous Green's function under a simultaneous exchange of spin, coordinate, time,  later extended to any orbital indices of the single-particle fermionic operators,  for a summary see e.g.  \cite{JL_AVB_OddwSC17}. 
The generalization allows for an extension of the possible symmetries of the superconducting order parameter to those that are both non-local and odd under the sign reversal of the relative time coordinate, or, equivalently, frequency. 
Originally proposed by Berezinskii \cite{Berezinskii74Rus, *Berezinskii74} as an odd-frequency spin-triplet $s$-wave order parameter for superfluid ${}^{3}$He, it was soon considered as a possible candidate in various superconducting systems, including disordered systems \cite{Kirkpatrick91}, heavy fermion \cite{Coleman93} and Kondo systems \cite{Coleman94, Coleman95, Fuseya2003}. 
More recently, it was considered in the context of topological materials \cite{Huang2015, Gnezdilov2019,tanaka2011symmetry}, such as Dirac semimetals \cite{SukhachovOddwDirac19}, but a conclusive evidence for a thermodynamically stable odd-frequency superconducting order parameter is still lacking.

Notwithstanding the issue of thermodynamic stability, various exotic properties in bulk superconducting materials are associated with enhanced pair correlations in the odd-frequency channel. A very promising route in generating odd-frequency pair correlations is in multiband (multiorbital, multichannel) systems \cite{BlackSchaffer2013, TriolaMultiOrb2020}, where, under fairly general conditions, such as band hybridization, odd-frequency interband pairing is generated even if the superconducting order parameter is strictly intraband and even-frequency. Candidates where such conditions are met were conceived in iron-pnictide superconductors \cite{Moreo2009, *Moreo2012, Youmans2018}, systems with disorder \cite{Komendova2015}, multi-channel Kondo systems \cite{Emery1992, CoxKondoEff98}, quantum spin Hall insulators on a honeycomb lattice \cite{Kuzmanovski2017}, and driven systems \cite{TriolaDriven2017}. In the presence of spin-orbit coupling, such as the case of Sr$_{2}$RuO$_{4}$, it leads to Kerr effect \cite{Komendova2017, Triola2018}.

SrTiO$_3$ (STO) is an example of multiband (multiorbital) superconductors, and hence  is an ideal candidate platform for observing the generation of odd-frequency pair correlations and associated features.
The bands crossing the Fermi level are due to the three $\mathrm{Ti}$ $t_{2g}$ orbitals, whose degeneracy is further lifted by spin-orbit coupling and the tetragonal crystal field due to antiferrodistortive rotations. 
The issue of the numerous  superconducting bands and gaps in STO is  a subject of a debate. 
The existence of multiple superconducting gaps is indicated by early tunneling measurement \cite{Binnig1980}, quantum oscillation measurement \cite{LinPRL2014}, and thermal conductivity data \cite{LinPRB2014}. 
These experimental data encouraged various theoretical studies about the multiband/multiorbital effects on the superconductivity in STO \cite{Fernandes2013,Edge2015Hc2,Trevisan2018,PhysRevB.100.094504}. 
On the other hand, recent tunneling experiment \cite{Swartz2018} and optical conductivity data \cite{Thiemann2018} suggest single superconducting gap in STO. 

Another intriguing feature of STO is the relationship between superconductivity and ferroelectric (FE) instability. 
The superconducting state in STO emerges in very low carrier density regime on the order of 10$^{17}$ cm$^{-3}$ \cite{Schooley1964,LinPRX2013,LinPRL2014,Bretz-Sullivan2019}, where the pairing mechanism cannot be captured by the Migdal-Eliashberg theory due to extreme retardation effects. 
Although various pairing interactions (e.g, plasmons \cite{Takada1980,Ruhman2016}, localized longitudinal optical modes \cite{Gorkov2016,Gorkov2017}) have been proposed to explain the persistence of superconductivity in the dilute density limit, the issue of the pairing mechanism has not yet been settled. 
On the other hand, STO is a quantum paraelectric (PE) in the vicinity of a FE quantum critical point \cite{Muller1979}, and which can undergo a FE transition under various chemical and physical actions such as isovalent substitution of Sr with Ca \cite{PhysRevLett.52.2289}, isotope substitution of oxygen \cite{PhysRevLett.82.3540}, strain \cite{PhysRevB.13.271}, and electric field \cite{PhysRevB.52.13159}. 
This FE instability motivated proposals of the pairing scenarios related to optical phonons mediating FE fluctuations \cite{Edge2015,Dunnett2018,Wolf2018,Arce-Gamboa2018,Kedem2018,vanderMarel2019, Gastiasoro2020,Sumita2020}. 
Indeed, enhancement of the superconducting transition temperature near a FE quantum critical point has been observed experimentally \cite{Stucky2016,Rischau2017,Tomioka2019,Ahadi2019,Herrera2019}. 
Furthermore, the coexistent phase of superconducting and FE-like orders, so-called FE superconducting state \cite{Kanasugi2018,PhysRevB.100.094504}, was recently observed in experiments \cite{Rischau2017,Russell2019}. 

In this work, we consider a three-orbital tight-binding model of STO in the vicinity of a FE critical point. 
Superconductivity is assumed to be due to some intraorbital spin-singlet isotropic order parameter. 
Under a FE transition, an additional Rashba-like spin-orbit coupling term appears proportional to the spontaneous FE order parameter that decreases the point group symmetry of the material from $D_{4h}$ to $C_{4v}$. 
We not only demonstrate that these ingredients guarantee the generation of odd-frequency pair correlations, but also classify them by combining the Berezinskii rule and the space group symmetry in the PE and FE phase of the system.
The classification table indicates that some components, although allowed by symmetry, are not dynamically generated by the terms of the normal state Hamiltonian that we take into account. 
Furthermore, we show that signatures of the odd-parity orbital hybridization, which is essential for the {\it ferroelectricity-induced} odd-frequency pair correlations, can be observed in the spectral functions and density of states (DOS). 

The rest of the article is organized as follows. 
In Sec. \ref{sec2}, we demonstrate a group theoretical classification of the multiorbital odd-frequency pair amplitudes based on the Berezinskii rule and space group symmetry of the system. 
As a specific example, we provide a classification table of the odd-frequency pair amplitudes in $t_{2g}$ electron systems such as STO. 
In Sec. \ref{sec3}, a three-orbital tight-binding model of bulk STO near a FE critical point is introduced. 
By using this model, we study the odd-frequency pair amplitudes in Sec. \ref{sec4}. 
It is shown that some components of the even-parity odd-frequency pair amplitudes are generated by the intrinsic LS coupling in the PE phase. 
Furthermore, we show that a FE transition induces the generation of additional odd-parity odd-frequency pair amplitudes and the enhancement of even-parity odd-frequency pair amplitudes. 
It is confirmed that the symmetry of the odd-frequency pair amplitudes is consistent with the results of the group theoretical classification in Sec. \ref{sec3}. 
In Sec. \ref{sec5}, we show that experimental signatures of the ferroelectricity-induced odd-parity hybridization can be observed in the orbital-resolved spectral functions and double-peak structure of DOS. 
Finally, a brief summary and conclusion are given in Sec. \ref{sec6}.

\section{\label{sec2} Symmetry of multiorbital odd-frequency pair correlations}
\subsection{\label{sec2-1} Fermionic symmetry}
We begin our discussion by classifying multiorbital superconducting states based on the fermionic symmetry, namely the Berezinskii rule \cite{AVB92,BlackSchaffer2013}. 
The pair amplitude, namely the anomalous Green's function, can be defined in a multiorbital system as  
\begin{equation}
\mathcal{F}_{ls,l's'}(\bm{k},\tau)=-\average{T_{\tau}c_{\bm{k},ls}(\tau)c_{-\bm{k},l's'}(0)}, 
\label{eq:F_tau}
\end{equation}
where $\tau$ is the imaginary time and $c_{\bm{k},ls}$ is the annihilation operator for an electron with momentum $\bm{k}$, orbital index $l$, and spin $s=\uparrow,\downarrow$. 
$T_{\tau}$ denotes the time-ordering operator for $\tau$. 
The Matsubara representation of Eq. (\ref{eq:F_tau}) is given by
\begin{equation}
\mathcal{F}_{ls,l's'}(\bm{k},i\omega_m)=\int_{0}^{\beta}d\tau\mathcal{F}_{ls,l's'}(\bm{k},\tau)e^{i\omega_m\tau} , 
\end{equation}
where $\omega_m=(2m+1)\pi\beta$ is a Matsubara frequency for the inverse temperature $\beta=1/T$.  
We here consider the spacial, time, spin, and orbital parities of Cooper pairs in a multiorbital system. 
The spacial inversion operation $\mathcal{P}$ is defined as 
\begin{equation}
\mathcal{P} \mathcal{F}_{ls,l's'}(\bm{k},i\omega_m) 
\equiv \mathcal{F}_{ls,l's'}(-\bm{k},i\omega_m) . 
\end{equation}
The time inversion operation $\mathcal{T}$, which changes sign of the relative time $\tau$ (not to be confused with the time reversal operation), acts on the Matsubara anomalous Green's function as 
\begin{equation}
\mathcal{T} \mathcal{F}_{ls,l's'}(\bm{k},i\omega_m) 
\equiv \mathcal{F}_{ls,l's'}(\bm{k},-i\omega_m). 
\end{equation}
The spin inversion operation $\mathcal{S}$ and the orbital index inversion operation $\mathcal{O}$,   simple swap operators permuting respective indices, are introduced as:
\begin{eqnarray}
\mathcal{S} \mathcal{F}_{ls,l's'}(\bm{k},i\omega_m) 
&\equiv& \mathcal{F}_{ls',l's}(\bm{k},i\omega_m), \\ 
\mathcal{O} \mathcal{F}_{ls,l's'}(\bm{k},i\omega_m) 
&\equiv& \mathcal{F}_{l's,ls'}(\bm{k},i\omega_m). 
\end{eqnarray}
The Fermi-Dirac statistics gives the sign change of the anomalous Green's function under the combined action of the spacial, time, spin, and orbital index inversion operations as
\begin{equation}
\mathcal{SPOT} \mathcal{F}_{ls,l's'}(\bm{k},i\omega_m)
= -\mathcal{F}_{ls,l's'}(\bm{k},i\omega_m) ,  
\label{eq:SPOT}
\end{equation}
which we write symbolically as $\mathcal{SPOT}=-1$. 
The full symmetries of the Cooper pair that satisfies Eq. (\ref{eq:SPOT}) are summarized in Table \ref{tab:classification_SPOT}. 

\begin{table}[tbp] 
\caption{\label{tab:classification_SPOT}
Symmetry properties of the anomalous Green's function under the spacial, time, spin, and orbital index inversion operations. 
}
\centering
\begin{ruledtabular}
{\renewcommand \arraystretch{1.3}
 \begin{tabular}{cccccc} 
	  Pairing & $\mathcal{S}$ & $\mathcal{P}$ & $\mathcal{O}$ & $\mathcal{T}$ & $\mathcal{SPOT}$  \\ \hline
  $-+++$ & $-1$ & $+1$ & $+1$ & $+1$ & $-1$ \\ 
  $---+$ & $-1$ & $-1$ & $-1$ & $+1$ & $-1$ \\ 
  $++-+$ & $+1$ & $+1$ & $-1$ & $+1$ & $-1$ \\
  $+-++$ & $+1$ & $-1$ & $+1$ & $+1$ & $-1$ \\ \hline
  $-+--$ & $-1$ & $+1$ & $-1$ & $-1$ & $-1$ \\ 
  $--+-$ & $-1$ & $-1$ & $+1$ & $-1$ & $-1$ \\ 
  $+++-$ & $+1$ & $+1$ & $+1$ & $-1$ & $-1$ \\
  $+---$ & $+1$ & $-1$ & $-1$ & $-1$ & $-1$ \\ 
\end{tabular}
}
\end{ruledtabular}
\end{table} 

Then, we perform the classification of the pair amplitude based on the results in Table \ref{tab:classification_SPOT}. 
First of all, the anomalous Green's function can be decomposed as 
\begin{equation}
\mathcal{F}_{ls,l's'}(k)=\left[\left(\psi_{ll'}(k)\sigma^{0}+\bm{d}_{ll'}(k)\cdot\bm{\sigma}\right)i\sigma^{y}\right]_{ss'}, 
\end{equation}
where $\sigma^0$ is a $2\times2$ identity matrix and $\bm{\sigma}=(\sigma^x, \sigma^y, \sigma^z)$ are the Pauli matrices. 
We here used the abbreviate notation $k=(\bm{k}, i\omega_m)$. 
The spin-singlet (triplet) pair amplitude $\psi_{ll'}(k)$ ($\bm{d}_{ll'}(k)$) is even (odd) under the spin inversion $\mathcal{S}$. 
Furthermore, we define the orbital-singlet and orbital-triplet pair amplitudes as 
\begin{align}
\psi_{ll'}^{\pm}(k)&=\frac{\psi_{ll'}(k) \pm \psi_{l'l}(k)}{2}, \label{eq:singlet} \\
\bm{d}_{ll'}^{\pm}(k)&=\frac{\bm{d}_{ll'}(k) \pm \bm{d}_{l'l}(k)}{2} . \label{eq:triplet}
\end{align}
The orbital-triplet pair amplitudes $\psi_{ll'}^{+}(k)$ and $\bm{d}_{ll'}^{+}(k)$ are even under the orbital index inversion $\mathcal{O}$. 
On the other hand, the orbital-singlet pair amplitudes $\psi_{ll'}^{-}(k)$ and $\bm{d}_{ll'}^{-}(k)$ are odd under the orbital index inversion $\mathcal{O}$. 

Using Eqs. (\ref{eq:singlet}) and (\ref{eq:triplet}), the odd-frequency pair amplitudes, that are odd under the time inversion $\mathcal{T}$, can be obtained as follows:
\begin{align}
\mathcal{F}_{ll'}^{-+--}&=\frac{(\psi_{ll'}^{-}+\mathcal{P}\psi_{ll'}^{-})-\mathcal{T}(\psi_{ll'}^{-}+\mathcal{P}\psi_{ll'}^{-})}{4} , \label{eq:F_-+--} \\
\mathcal{F}_{ll'}^{--+-}&=\frac{(\psi_{ll'}^{+}-\mathcal{P}\psi_{ll'}^{+})-\mathcal{T}(\psi_{ll'}^{+}-\mathcal{P}\psi_{ll'}^{+})}{4} , \label{eq:F_--+-} \\
\bm{\mathcal{F}}_{ll'}^{+++-}&=\frac{(\bm{d}_{ll'}^{+}+\mathcal{P}\bm{d}_{ll'}^{+})-\mathcal{T}(\bm{d}_{ll'}^{+}+\mathcal{P}\bm{d}_{ll'}^{+})}{4} , \label{eq:F_+++-} \\
\bm{\mathcal{F}}_{ll'}^{+---}&=\frac{(\bm{d}_{ll'}^{-}-\mathcal{P}\bm{d}_{ll'}^{-})-\mathcal{T}(\bm{d}_{ll'}^{-}-\mathcal{P}\bm{d}_{ll'}^{-})}{4} , \label{eq:F_+---}
\end{align}
where we suppressed the $k$-dependence for brevity. 
The spin-triplet pair amplitudes are described by using a vector notation $\bm{\mathcal{F}}_{ll'}=\mathcal{F}_{ll',x}\hat{\bm{x}}+\mathcal{F}_{ll',y}\hat{\bm{y}}+\mathcal{F}_{ll',z}\hat{\bm{z}}$. 
The above odd-frequency pair amplitudes satisfy the Berezinskii rule (i.e., $\mathcal{SPOT}=-1$).

\subsection{\label{sec2-2} Space group symmetry}
In the previous section, we classified a multiorbital odd-frequency superconducting state based on the fermionic symmetry of the Cooper pair. 
 Superconducting state can also lower point group symmetry of the state and any pairing state can be classified by the space group symmetry of the crystal structure \cite{Sigrist-Ueda}. 
Here, we provide a classification of multiorbital superconducting states based on the space group symmetry by considering the transformation of the Bloch wave function. 

A creation operator of a Bloch state with orbital index $l$ and spin $s$ can be defined as 
\begin{align}
c_{\bm{k},ls}^{\dag}=\sum_{\bm{R}}c_{ls}^{\dag}(\bm{R}) e^{-i\bm{k}\cdot\bm{R}}  , \label{eq:Bloch}
\end{align}
where $\bm{R}$ represents the position for the unit cell (lattice vector).
Using Eq. (\ref{eq:Bloch}), the creation operator is transformed by a space group operation $g=\{p|\bm{a}\}$ as follows: 
\begin{eqnarray}
g c_{\bm{k},ls}^{\dag} g^{-1} 
=&& \sum_{\bm{R}} g c_{ls}^{\dag}(\bm{R}) g^{-1} e^{-i\bm{k}\cdot\bm{R}} \nonumber\\
=&& \sum_{\bm{R}} e^{-i\bm{k}\cdot\bm{R}} \sum_{l',s'} c_{l's'}^{\dag}(p\bm{R}+\bm{a}) \nonumber\\ 
&&\times D_{l'l}^{({\rm orb})}(p) D_{s's}^{(1/2)}(p) ,
\label{eq:Bloch_G_def}
\end{eqnarray}
where $D^{(1/2)}(p)$ and $D^{({\rm orb})}(p)$ are the representation matrices for a point group operation $p$ in spin and orbital space, respectively. 
By defining $\bm{R}'\equiv p\bm{R}+\bm{a}$, Eq. (\ref{eq:Bloch_G_def}) is rewritten as
\begin{eqnarray}
g c_{\bm{k},ls}^{\dag} g^{-1}  
=&& e^{ip\bm{k}\cdot\bm{a}} \sum_{l',s'} \left( \sum_{\bm{R}'} e^{-ip\bm{k}\cdot\bm{R}'}c_{l's'}^{\dag}(\bm{R}') \right) \nonumber\\
&&\times D_{l'l}^{({\rm orb})}(p) D_{s's}^{(1/2)}(p), \nonumber\\
=&& e^{ip\bm{k}\cdot\bm{a}} \sum_{l',s'} c_{p\bm{k},l's'}^{\dag} D_{l'l}^{({\rm orb})}(p) D_{s's}^{(1/2)}(p). \label{eq:Bloch_G}
\end{eqnarray}
From Eq. (\ref{eq:Bloch_G}), the pair amplitude is transformed by a space group operation $g=\{p|\bm{a}\}$ as 
\begin{eqnarray}
g \mathcal{F}_{ls,l's'}(\bm{k},i\omega_m) g^{-1} 
=&& \sum_{\lambda,\lambda'}\sum_{\sigma,\sigma'} \mathcal{F}_{\lambda\sigma,\lambda'\sigma'}(p\bm{k},i\omega_m) \mathcal{D}^{\Gamma}(g) \nonumber\\
&&\times\mathcal{D}_{\lambda\lambda', l l'}^{({\rm orb})}(p) \mathcal{D}_{\sigma\sigma', s s'}^{(1/2)}(p), \label{eq:gap_Gtrans} 
\end{eqnarray}
where the corresponding representation matrices are
\begin{align}
\mathcal{D}_{\lambda \lambda', l l'}^{({\rm orb})}(p)
&= D_{\lambda l}^{({\rm orb})}(p) D_{\lambda' l'}^{({\rm orb})}(p), \\
\mathcal{D}_{\sigma \sigma', s s'}^{(1/2)}(p)
&= D_{\sigma s}^{(1/2)}(p) D_{\sigma' s'}^{(1/2)}(p) , 
\end{align}
and $\mathcal{D}^{\Gamma}(g)$ is the representation matrix of the $\Gamma$ irreducible representation (IR) of the gap function. 
Note that the frequency dependence of the pair amplitude is not changed by the space group operations. 
By using Eq. (\ref{eq:gap_Gtrans}), we can classify the $\bm{k}$-dependence of pair amplitudes under a given space group symmetry. 

\subsection{\label{sec2-3} Application to polar $t_{2g}$ electron systems}
In this section, we perform the classification of the multiorbital odd-frequency pair amplitude based on both fermionic and space group symmetries. 
As far as we know, we here provide a first example of group theoretical classification for the  multiorbital odd-frequency superconductivity, although that for the multiorbital even-frequency superconductivity \cite{PhysRevB.94.174513} and single-orbital odd-frequency superconductivity \cite{PhysRevB.97.024507} has been done in previous works. 
Note that we can classify the even-frequency pair amplitude in the same manner (see Appendix \ref{sec:even-w}).  

As a demonstration, we here consider a time-reversal symmetric $t_{2g}$ electron system in tetragonal $D_{4h}$ or $C_{4v}$ point group. 
To simplify the discussion, we assume $s$-wave superconducting states which belong to trivial IR (i.e., $\mathcal{D}^{\Gamma}(g)=1$). 
Then, according to Eq. (\ref{eq:gap_Gtrans}), it is sufficient to classify the pair amplitudes based on the point group symmetry, not space group symmetry. 
Extension to anisotropic superconducting states that transform in accordance with nontrivial IRs is straightforward. 
The basis functions for $\bm{k}$-dependence of the odd-frequency pair amplitudes, that are determined by using Eq. (\ref{eq:gap_Gtrans}), are listed in Table \ref{tab:classification_oddw}. 
When the anomalous Green's function $\mathcal{F}_{ls,l's'}(k)$ belongs to the trivial IR of $D_{4h}$ point group, that is $A_{1g}$ IR, only the even-parity ($\mathcal{P}=+1$) pair amplitudes $\mathcal{F}^{-+--}$ and $\bm{\mathcal{F}}^{+++-}$ in Table \ref{tab:classification_oddw} are allowed. 
Then, the odd-parity ($\mathcal{P}=-1$) pair amplitudes should vanish because they do not belong to $A_{1g}$ IR due to the sign change under spacial inversion operation $\mathcal{P}$. 
On the other hand, the odd-parity pair amplitudes may be trivial under all symmetry operations in $C_{4v}$ point group, since $D_{4h}$ point group can be decomposed as $D_{4h}=C_{i}\otimes C_{4v}$. 
Therefore, the odd-parity pair amplitudes $\mathcal{F}^{--+-}$ and $\bm{\mathcal{F}}^{+---}$ belonging to $A_{2u}$ IR in $D_{4h}$ point group are listed in Table \ref{tab:classification_oddw}, because $A_{2u}$ is reduced to the trivial $A_{1}$ IR in $C_{4v}$ point group. 
In the microscopic analysis of a three-orbital model (Secs. \ref{sec3} and \ref{sec4}), we obtain pair amplitudes compatible with Table \ref{tab:classification_oddw}. 

The odd-parity $A_{2u}$ pair amplitudes are symmetrically forbidden in centrosymmetric $D_{4h}$ point group, when the superconductivity belongs to the trivial IR as expected in STO. 
In order to induce the odd-parity $A_{2u}$ pair amplitudes, a spacial inversion symmetry breaking, which descends the crystallographic point group from $D_{4h}$ to $C_{4v}$, is necessary.
This symmetry lowering from $D_{4h}$ to $C_{4v}$ can be realized by a FE phase transition, such as a polar inversion symmetry breaking that actually occurs in STO \cite{PhysRevLett.52.2289, PhysRevLett.82.3540,PhysRevB.13.271, PhysRevB.52.13159}.

\begin{table*}[htbp] 
\caption{\label{tab:classification_oddw}
Basis functions for $\bm{k}$-dependence of the odd-frequency pair amplitude in a $t_{2g}$ electron system under $D_{4h}$ point group symmetry. 
Basis functions of $A_{1g}$ and $A_{2u}$ IRs are listed, because the superconducting gap function is assumed to belong the trivial IR of $D_{4h}$ or $C_{4v}$. 
The orbital index $l=1,2,3$ refers to $d_{yz}, d_{xz}, d_{xy}$ orbitals, respectively. $\phi_{j-}^{\Gamma}$ and $\eta_{j-}^{\Gamma}$ ($\Gamma=A_{1g}, A_{2u}$ and $j=1,2,\cdots$) are arbitrary constants. 
}
\centering
\begin{ruledtabular}
{\renewcommand \arraystretch{1.4}
\begin{tabular}{ccc} 
	$\mathcal{F}_{ll'}^{-+--}$  & $\bm{k}$-dependence of $\mathcal{F}_{ll'}^{-+--}(\bm{k},i\omega_m)$ & IR  \\ \hline 
	$\mathcal{F}_{12}^{-+--}$
			& $\phi^{A_{1g}}_{1-}k_xk_y(k_x^2-k_y^2)$ & \multirow{3}{*}{$A_{1g}$} \\ 
   $\mathcal{F}_{13}^{-+--}$
			& $\phi^{A_{1g}}_{2-}k_xk_z$  \\
   $\mathcal{F}_{23}^{-+--}$
			& $\phi^{A_{1g}}_{2-}k_yk_z$  \\ \hline\hline 
   $\mathcal{F}_{ll'}^{--+-}$ & $\bm{k}$-dependence of $\mathcal{F}_{ll'}^{--+-}(\bm{k},i\omega_m)$ & IR  \\ \hline
  $\mathcal{F}_{11}^{--+-}$
     & $\phi^{A_{2u}}_{1-}k_z$  &  \multirow{6}{*}{$A_{2u}$} \\  
  $\mathcal{F}_{22}^{--+-}$
     & $\phi^{A_{2u}}_{1-}k_z$ \\
  $\mathcal{F}_{33}^{--+-}$
     & $\phi^{A_{2u}}_{2-}k_z$  \\
  $\mathcal{F}_{12}^{--+-}$
     & $\phi^{A_{2u}}_{3-}k_xk_yk_z$ \\
  $\mathcal{F}_{13}^{--+-}$
     & $\phi^{A_{2u}}_{4-}k_x$ \\
  $\mathcal{F}_{23}^{--+-}$
     & $\phi^{A_{2u}}_{4-}k_y$ \\ \hline\hline
   $\bm{\mathcal{F}}_{ll'}^{+++-}$ & $\bm{k}$-dependence of $\bm{\mathcal{F}}_{ll'}^{+++-}(\bm{k},i\omega_m)$ & IR  \\ \hline
  $\bm{\mathcal{F}}_{11}^{+++-}$ 
      & $\eta^{A_{1g}}_{1-}k_yk_z\hat{\bm{x}}+\eta^{A_{1g}}_{2-}k_xk_z\hat{\bm{y}}+\eta^{A_{1g}}_{3-}k_xk_y\hat{\bm{z}}$ & \multirow{6}{*}{$A_{1g}$} \\ 
  $\bm{\mathcal{F}}_{22}^{+++-}$ 
      & $-\eta^{A_{1g}}_{2-}k_yk_z\hat{\bm{x}}-\eta^{A_{1g}}_{1-}k_xk_z\hat{\bm{y}}-\eta^{A_{1g}}_{3-}k_xk_y\hat{\bm{z}}$ \\ 
  $\bm{\mathcal{F}}_{33}^{+++-}$ 
      & $\eta^{A_{1g}}_{4-}(k_yk_z\hat{\bm{x}}-k_xk_z\hat{\bm{y}})+\eta^{A_{1g}}_{5-}k_xk_y(k_x^2-k_y^2)\hat{\bm{z}}$ \\ 
  $\bm{\mathcal{F}}_{12}^{+++-}$ 
      & $\eta^{A_{1g}}_{6-}(k_xk_z\hat{\bm{x}}-k_yk_z\hat{\bm{y}})+\eta^{A_{1g}}_{7-}(k_x^2-k_y^2)\hat{\bm{z}}$ \\ 
  $\bm{\mathcal{F}}_{13}^{+++-}$ 
      & $\eta^{A_{1g}}_{8-}k_xk_y\hat{\bm{x}}+(\eta^{A_{1g}}_{9-}+\eta^{A_{1g}}_{10-}k_x^2+\eta^{A_{1g}}_{11-}k_y^2+\eta^{A_{1g}}_{12-}k_z^2)\hat{\bm{y}}+\eta^{A_{1g}}_{13-}k_yk_z\hat{\bm{z}}$ \\ 
  $\bm{\mathcal{F}}_{23}^{+++-}$ 
      & $-(\eta^{A_{1g}}_{9-}+\eta^{A_{1g}}_{11-}k_x^2+\eta^{A_{1g}}_{10-}k_y^2+\eta^{A_{1g}}_{12-}k_z^2)\hat{\bm{x}}-\eta^{A_{1g}}_{8-}k_xk_y\hat{\bm{y}}-\eta^{A_{1g}}_{13-}k_xk_z\hat{\bm{z}}$  \\ \hline\hline
   $\bm{\mathcal{F}}_{ll'}^{+---}$ & $\bm{k}$-dependence of $\bm{\mathcal{F}}_{ll'}^{+---}(\bm{k},i\omega_m)$ & IR  \\ \hline
  $\bm{\mathcal{F}}_{12}^{+---}$ 
   & $\eta^{A_{2u}}_{1-}(k_x\hat{\bm{x}}+k_y\hat{\bm{y}})+\eta^{A_{2u}}_{2-}k_z\hat{\bm{z}}$ & \multirow{3}{*}{$A_{2u}$} \\ 
  $\bm{\mathcal{F}}_{13}^{+---}$ 
   & $\eta^{A_{2u}}_{3-}k_xk_yk_z\hat{\bm{x}}+\eta^{A_{2u}}_{4-}k_z\hat{\bm{y}}+\eta^{A_{2u}}_{5-}k_y\hat{\bm{z}}$ \\ 
  $\bm{\mathcal{F}}_{23}^{+---}$ 
   & $-\eta^{A_{2u}}_{4-}k_z\hat{\bm{x}}-\eta^{A_{2u}}_{3-}k_xk_yk_z\hat{\bm{y}}-\eta^{A_{2u}}_{5-}k_x\hat{\bm{z}}$ \\ 
\end{tabular}
}
\end{ruledtabular}
\end{table*} 

\section{\label{sec3} Model and method}
In this section, we introduce an electron-lattice coupled multiorbital model for bulk STO.  

\subsection{\label{sec3-1} Three-orbital model}
First, we introduce a three-orbital tight-binding model for $t_{2g}$ electrons in bulk STO as follows:
\begin{align}
\mathcal{H} &= \mathcal{H}_{\rm kin} +\mathcal{H}_{\rm LS}+\mathcal{H}_{\rm pol}+\mathcal{H}_{\rm pair}, \\
\mathcal{H}_{\rm kin} &= 
\sum_{{\bm k},l,s} \xi_l({\bm k})  c_{{\bm k}l s}^{\dag}c_{{\bm k} l s} , \label{eq:H_kin} \\
\mathcal{H}_{\rm LS} &= \frac{\lambda}{2} \sum_{\bm{k},l,l',s,s'} \sum_{\mu=x,y,z}\hat{l}_{ll'}^{\mu} \otimes \hat{\sigma}_{ss'}^{\mu} c_{\bm{k}ls}^{\dag} c_{\bm{k}l's'}, \label{eq:H_LS} \\
\mathcal{H}_{\rm pol} &= \sum_{{\bm k},s} \sum_{l=1,2} \left[\zeta_{l}(\bm{k}) c_{\bm{k} ls}^{\dag} c_{\bm{k} 3s} + \rm{H.c.} \right] , \label{eq:H_pol} \\
\mathcal{H}_{\rm pair} &= -\frac{V_{s}}{N} 
\sum_{{\bm k},{\bm k}',l}
c_{{\bm k} l \uparrow}^{\dag} c_{-{\bm k} l \downarrow}^{\dag}
c_{-{\bm k}' l \downarrow} c_{{\bm k}' l \uparrow}, 
\label{eq:H_pair} 
\end{align}
where $d_{yz}, d_{xz}, d_{xy}$ orbitals are denoted by the orbital index $l=1,2,3$, respectively. 
The tight-binding forms are given by
\begin{eqnarray}
\xi_{1}({\bm k})=&&-2 t_1 \left(\cos k_y + \cos k_z \right)
-2 t_2 \cos k_x \nonumber\\ && -4 t_3 \cos k_y \cos k_z -\mu, \\
\xi_{2}({\bm k})=&& -2 t_1 \left(\cos k_x + \cos k_z\right)
-2 t_2 \cos k_y \nonumber\\ && -4 t_3 \cos k_x \cos k_z -\mu, \\
\xi_{3}({\bm k})=&&-2 t_1 \left(\cos k_x + \cos k_y\right)
-2 t_2 \cos k_z \nonumber\\ && -4 t_3 \cos k_x \cos k_y + \Delta_{\rm T} -\mu, \\
\zeta_{1,2}(\bm{k}) =&& 2i\gamma\sin k_{x,y} .
\end{eqnarray}
The first term $\mathcal{H}_{\rm kin}$ is the kinetic-energy term of $t_{2g}$ orbitals. 
The single electron kinetic energy $\xi_{l}({\bm k})$ includes the chemical potential $\mu$ and the tetragonal crystal field $\Delta_{\rm T}$ arising from the antiferrodistortive structural transition. 
The chemical potential $\mu$ is determined to fix the carrier density per Ti site as $n$. 
The second term $\mathcal{H}_{\rm LS}$ represents the atomic LS coupling of Ti ions. 
The orbital angular momentum operator $\hat{\bm{l}}=(\hat{l}^x, \hat{l}^y, \hat{l}^z)$ is defined as 
\begin{align}
\hat{\bm{l}}=\left(
\begin{pmatrix}
 0 & 0 & 0 \\
 0 & 0 & i \\
 0 & -i & 0
\end{pmatrix}, 
\begin{pmatrix}
 0 & 0 & -i \\
 0 & 0 & 0 \\
 i & 0 & 0
\end{pmatrix}, 
\begin{pmatrix}
 0 & i & 0 \\
 -i & 0 & 0 \\
 0 & 0 & 0
\end{pmatrix} 
\right) , 
\end{align}
in the orbital basis $(d_{yz},d_{xz}, d_{xy})$. 
Here, $\hat{\bm{l}}$ is the projection of the $L=2$ angular momentum operator onto the $t_{2g}$ orbital subspace. 
The band structure of bulk STO is reproduced by $\mathcal{H}_{\rm kin}+\mathcal{H}_{\rm LS}$ with the parameter set $(t_1, t_2, t_3, \Delta_{\rm T}, \lambda)=(1, 0.11, 0.27, 0.03, 0.12)$ \cite{hirayama2012ab,PhysRevB.86.125121,PhysRevB.87.161102,PhysRevB.100.094504}, where the unit of energy is chosen as $t_1=1$. 
The third term $\mathcal{H}_{\rm pol}$ is an intersite odd-parity orbital hybridization term. 
Since $\mathcal{H}_{\rm pol}$ belongs to $A_{2u}$ IR of $D_{4h}$ point group, 
it is allowed only in the FE phase with broken mirror symmetry along the $[001]$ axis. 
Then, the hopping integral $\gamma$ can be treated as a FE order parameter $P_z$ which characterize the polar inversion symmetry breaking along the [001] axis \cite{PhysRevB.100.094504} (i.e., $\gamma \propto P_z$). 
Combination of $\mathcal{H}_{\rm LS}$ and $\mathcal{H}_{\rm pol}$ leads to the Rashba-type spin-orbit splitting of the band structure in the FE phase. 
We study superconductivity by adopting the BCS-type static attractive interaction $\mathcal{H}_{\rm pair}$. 
For simplicity, we assume  momentum-independent intraorbital pairing interaction. 
$N$ is the number of Ti sites, and $V_s$ is the pairing interaction strength. 

In this study, we apply the mean-field approximation to $\mathcal{H}_{\rm pair}$ as
\begin{align}
\mathcal{H}_{\rm pair} 
&\approx \sum_{{\bm k},l} \left(
\Delta_{l} c_{{\bm k}l \uparrow}^{\dag} c_{-{\bm k} l \downarrow}^{\dag}
+ \rm{H.c.} \right) 
+\frac{N}{V_s}\sum_{l}|\Delta_{l}|^2, 
\end{align}
by introducing the orbital-dependent superconducting order parameter, 
\begin{equation}
\Delta_{l} = - \frac{V_{s}}{N} \sum_{{\bm k}}
\average{c_{-{\bm k}l\downarrow} c_{{\bm k}l\uparrow}} .
\label{eq:gap_STO}
\end{equation}
Note that $\Delta_{1}=\Delta_{2}$ under 4-fold rotational symmetry. 
By introducing the multiorbital Nambu spinors
\begin{align}
\hat{\Psi}_{\bm{k}}^{\dag} &= 
( \hat{\psi}_{\bm{k}}^{\dag}, \hat{\psi}_{-\bm{k}}^{\rm T} ), \\
\hat{\psi}_{\bm{k}}^\dag &= ( 
c_{{\bm k}1\uparrow}^{\dag}, c_{{\bm k}2\uparrow}^{\dag}, 
c_{{\bm k}3\uparrow}^{\dag}, c_{{\bm k}1\downarrow}^{\dag}, 
c_{{\bm k}2\downarrow}^{\dag}, c_{{\bm k}3\downarrow}^{\dag}), 
\end{align}
we obtain the matrix representation of the mean-field Hamiltonian as
\begin{align}
\mathcal{H} = \frac{1}{2}\sum_{{\bm k}}
\hat{\Psi}_{\bm k}^{\dag} \hat{\mathcal{H}}(\vector{k}) \hat{\Psi}_{\bm k} 
 + \frac{N}{V_s} \sum_{l} |\Delta_{l}|^2 + \sum_{\bm{k},l} \xi_l(\bm{k}).
\label{eq:H_mf_mat}
\end{align}
The Bogoliubov-de Gennes (BdG) Hamiltonian $\hat{\mathcal{H}}(\vector{k})$ is described as 
\begin{equation}
 \hat{\mathcal{H}}(\vector{k})= 
    \begin{pmatrix}
       \hat{\mathcal{H}}_{0}(\vector{k}) & \hat{\Delta}  \\ 
      \hat{\Delta}^{\dag} &  -\hat{\mathcal{H}}_{0}^{\rm T}(-\vector{k})
    \end{pmatrix} , 
\label{eq:BdG}
\end{equation}
where the normal state part $\hat{\mathcal{H}}_{0}(\bm{k})$ and pairing part $\hat{\Delta}$ are given by
\begin{widetext}
\begin{align}
\hat{\mathcal{H}}_{0}(\bm{k}) = 
\begin{pmatrix}
 \xi_{1}(\bm{k}) &i\lambda/2 & 
 \zeta_{1} (\bm{k}) & 0 & 
 0 & -\lambda/2 \\ 
 -i\lambda/2 & \xi_{2}(\bm{k}) & 
 \zeta_{2}(\bm{k}) & 0 & 
 0 & i\lambda/2 \\ 
 \zeta_{1}^*(\bm{k}) & \zeta_{2}^*(\bm{k}) & 
 \xi_{3}(\bm{k}) & \lambda/2 & 
 -i\lambda/2 & 0 \\
 0 & 0 & 
 \lambda/2 & \xi_{1}(\bm{k}) & 
 -i\lambda/2 & \zeta_{1}(\bm{k}) \\ 
 0 & 0 & 
 i\lambda/2 & i\lambda/2 & 
 \xi_{2}(\bm{k}) & \zeta_{2}(\bm{k}) \\ 
 -\lambda/2 & -i\lambda/2 & 
 0 & \zeta_{1}^*(\bm{k}) & 
 \zeta_{2}^*(\bm{k}) & \xi_{3}(\bm{k})
\end{pmatrix} , \quad
\hat{\Delta} = 
\begin{pmatrix}
 0 & 0 & 0 & \Delta_{1} & 0 & 0 \\ 
 0 & 0 & 0 & 0 & \Delta_{2} & 0 \\ 
 0 & 0 & 0 & 0 & 0 & \Delta_{3} \\ 
 -\Delta_{1} & 0 & 0 & 0 & 0 & 0 \\ 
 0 & -\Delta_{2} & 0 & 0 & 0 & 0 \\ 
 0 & 0 & -\Delta_{3} & 0 & 0 & 0 
\end{pmatrix} .
\label{eq:H_BdG}
\end{align}
\end{widetext}

 The BdG Hamiltonian $\hat{\mathcal{H}}(\vector{k})$ is diagonalized by carrying out the Bogoliubov transformation as 
\begin{align}
c_{{\bm k}, l\sigma} = \sum_{\nu,\tau} \left( u_{{\bm k},l\sigma}^{(\nu\tau)} \alpha_{{\bm k},\nu\tau} + v_{-{\bm k},l\sigma}^{(\nu\tau)*} \alpha_{-{\bm k},\nu\tau}^{\dag} \right) ,
\end{align}
where $\alpha_{{\bm k},\nu\tau}$ is the annihilation operator for a Bogoliubov quasiparticle with momentum $\bm{k}$, pseudoorbital $\nu=1,2,3$, and pseudospin $\tau=\uparrow,\downarrow$.  
Then, Eq. (\ref{eq:H_mf_mat}) is rewritten as
\begin{eqnarray}
\mathcal{H} =&&
\sum_{{\bm k},\nu,\tau}  E_{\nu\tau}(\bm{k}) \left( \alpha_{{\bm k},\nu\tau}^{\dag} \alpha_{{\bm k},\nu\tau} - \frac{1}{2} \right) \nonumber\\
&&+ \frac{N}{V_s} \sum_{l} |\Delta_{l}|^2 +\sum_{\bm{k},l} \xi_l(\bm{k}) ,
\end{eqnarray}
and hence the electronic free energy per Ti site is obtained as 
\begin{eqnarray}
\Omega_{\rm ele}[\bm{\Delta},P_z] =&& -\frac{1}{N\beta} \sum_{{\bm k},\nu,\tau} \left[ \ln \left(1+e^{-\beta E_{\nu \tau}(\bm{k})} \right) + \frac{\beta E_{\nu \tau}(\bm{k})}{2} \right] \nonumber\\ &&+ \frac{1}{V_s} \sum_{l} |\Delta_{l}|^2 + \frac{1}{N} \sum_{\bm{k},l} \xi_l(\bm{k}) + \mu n .
\label{eq:Helmholtz_STO}
\end{eqnarray}
Here, we used an abbreviate notation $\bm{\Delta}=(\Delta_{1}, \Delta_{2}, \Delta_{3})$. 
The last term of Eq. (\ref{eq:Helmholtz_STO}) is necessary 
since the carrier density per Ti site is fixed as $n$ instead of fixing the chemical potential $\mu$. 
Using $\bm{\Delta}$ obtained by solving Eq. (\ref{eq:gap_STO}), we calculate the electronic part of the free energy $\Omega_{\rm ele}[\bm{\Delta},P_z]$ from Eq. (\ref{eq:Helmholtz_STO}).

\subsection{\label{sec3-2} Polar lattice distortion}
Next, we introduce the Landau free energy arising from the polar lattice distortion as
\begin{equation}
\Omega_{\rm lat} [P_z] = \frac{1}{2} \kappa_{2} P_z^2 + \frac{1}{4} \kappa_{4} P_z^4 + \frac{1}{6} \kappa_{6} P_z^6, 
\end{equation}
where $\kappa_2$, $\kappa_4$, and $\kappa_6$ are the lattice parameters which describe the elasticity of the lattice \cite{PhysRevB.100.094504}. 
The temperature dependence of lattice parameters is neglected in accordance with the fact that the dielectric constant in bulk STO is almost constant at low temperatures \cite{Muller1979}. 

Then, the total free energy including the contributions of both electrons and lattice is given by
\begin{equation}
\Omega[\bm{\Delta},P_z] = \Omega_{\rm ele}[\bm{\Delta},P_z] + \Omega_{\rm lat}[P_z] . 
\end{equation}
The thermodynamically stable state is determined by minimizing the free energy $\Omega[\bm{\Delta},P_z]$ with respect to $\bm{\Delta}$ and $P_z$ \cite{Kanasugi2018,PhysRevB.100.094504}. 
The values of $\kappa_4$ and $\kappa_6$ are determined to cut off the FE order parameter $P_z\propto\gamma$ in a realistic regime $\gamma/t_1\lesssim 0.20$ \cite{PhysRevB.87.161102,PhysRevB.100.094504}. 
The lattice parameter $\kappa_2$ is a control parameter for the ferroelectricity, and we determined its value so as to realize a PE normal state near a FE critical point. 
Then, we consider the superconductivity in bulk STO which is tuned toward the vicinity of the FE critical point by isovalent substitution of Sr with Ca \cite{PhysRevLett.52.2289}, isotope substitution of oxygen \cite{PhysRevLett.82.3540}, strain \cite{PhysRevB.13.271}, electric field \cite{PhysRevB.52.13159}, and by any other means. 

\subsection{\label{sec3-3} Numerical setup}
To study the interplay of odd-frequency superconductivity and FE transition in bulk STO, we perform numerical calculations in a dilute carrier density regime in which both the superconductivity and ferroelectricity can be stabilized simultaneously owing to the Lifshitz transition \cite{PhysRevB.100.094504}.  

The carrier density is fixed as $n=5.0\times10^{-5}$ which corresponds to $8.40\times10^{17}$ cm$^{-3}$ by adopting the lattice constant of STO ($a=3.905$  {\AA}) as the unit of length. 
The pairing interaction is chosen to be $V_{\rm s}/t_1=0.28$ which gives the superconducting transition temperature $T_{\rm c}=0.00098t_1\sim 3.0$ K at $\gamma=0$. 
The value of $T_{\rm c}$ is set to be larger than the realistic transition temperature ($\sim$ 0.3 K) to reduce the cost of the numerical calculation. 
The cutoff lattice parameters are chosen as $(\kappa_4, \kappa_6)=(0,0.5)$. 
Then, the coexistent phase of superconductivity and ferroelectricity is stabilized when the lattice parameter $\kappa_2$ is set to be smaller than a critical value. 
The value of $\kappa_2$ and obtained results of the superconducting and FE order parameters at almost zero temperature ($T/t_1=1.0\times10^{-10}$) are summarized in Table \ref{tab:FESC}. 

\begin{table}[tbp] 
\caption{\label{tab:FESC}
Model parameters for the numerical calculations. 
The superconducting order parameter $\Delta_{l}$ and odd-parity hoping integral $\gamma$ are determined by minimizing the free energy at $T/t_1=1.0\times10^{-10}$. 
}
\centering
\begin{ruledtabular}
{\renewcommand \arraystretch{1.3}
 \begin{tabular}{ccc} 
    & PE phase & FE phase \\ \hline
 $\kappa_2$ & $8.00\times10^{-5}$ & $6.75\times10^{-5}$ \\
 $\gamma$ & 0 & 0.105 \\
 $\Delta_{1,2}$ & 0.00160 & 0.00277 \\
 $\Delta_{3}$ & 0.000268 & 0.00138  
\end{tabular}
}
\end{ruledtabular}
\end{table} 

\section{\label{sec4} Odd-frequency pair correlations}
On the basis of numerical solution of the electron-lattice coupled three-orbital model, we elucidate odd-frequency pair amplitudes in bulk STO. 
Using the BdG Hamiltonian matrix in Eq. (\ref{eq:H_BdG}), we obtain the full Matsubara Green's function matrix as 
\begin{equation}
\hat{\mathcal{G}}(\bm{k},i\omega_m)=\left[ i\omega_m \hat{\bm{1}} - \hat{\mathcal{H}}(\bm{k}) \right]^{-1} , 
\end{equation}
where $\hat{\bm{1}}$ is a $12\times12$ identity matrix. 
The matrix structure of $\hat{\mathcal{G}}(k)$ is described as 
\begin{align}
\hat{\mathcal{G}}(k)&=
\begin{pmatrix}
 \hat{\mathcal{G}}^{0}(k) & \hat{\mathcal{F}}(k) \\ 
 \hat{\bar{\mathcal{F}}}(k) & \hat{\bar{\mathcal{G}}}^{0}(k) 
\label{eq:Matsubara_G}
\end{pmatrix} .
\end{align}
The $6\times6$ submatrix $\hat{\mathcal{G}}^{0}(k)$ and $\hat{\mathcal{F}}(k)$ are the normal and anomalous Green's functions, respectively. 
$\hat{\bar{\mathcal{G}}}^{0}(k)$ ($\hat{\bar{\mathcal{F}}}(k)$) is the particle-hole counterpart of $\hat{\mathcal{G}}^{0}(k)$ ($\hat{\mathcal{F}}(k)$). 
By applying Eqs. (\ref{eq:F_-+--})-(\ref{eq:F_+---}) to the matrix elements of $\hat{\mathcal{F}}(k)$, we calculate the odd-frequency pair amplitudes in bulk STO.

\subsection{\label{sec4-1} Paraelectric phase}
\begin{figure}[tbp]
\centering
   \includegraphics[width=86mm,clip]{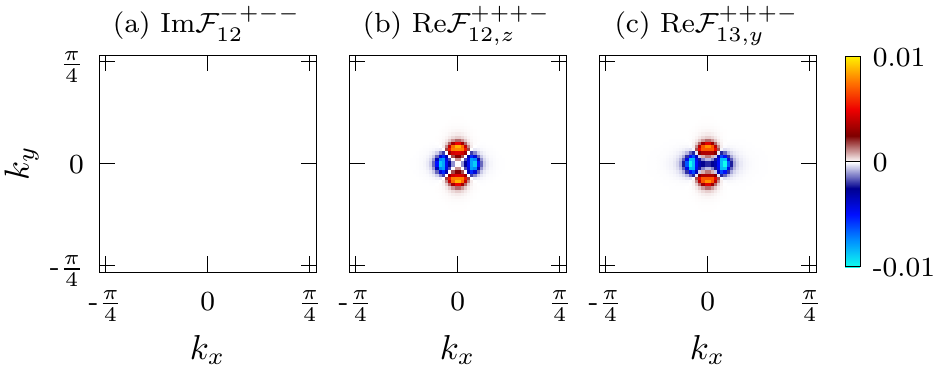}
   \caption{$\bm{k}$-dependence of the even-parity odd-frequency pair amplitudes (a) $\mathrm{Im}\mathcal{F}_{12}^{-+--}(k)$, (b) $\mathrm{Re}\mathcal{F}_{12,z}^{+++-}(k)$, and (c) $\mathrm{Re}\mathcal{F}_{13,y}^{+++-}(k)$ at $k_z=0$ in the PE phase. The Matsubara frequency $\omega_m$ is set to be 1 meV, and the values of the pair amplitudes are normalized by $\mathcal{F}^{\rm BCS}_{\rm max}(\omega_m)=0.237808$ meV$^{-1}$. 
 \label{fig:Fodd_PE} }
\end{figure}

First, we consider the odd-frequency pairing in the PE phase with $D_{4h}$ point group symmetry. 
Since we assumed the $s$-wave ($A_{1g}$) pairing [Eq. (\ref{eq:H_pair})], the odd-frequency pair amplitudes should belong to $A_{1g}$ IR in the PE phase. 
Then, the even-parity $A_{1g}$ odd-frequency pair amplitudes can be finite. 
The orbital hybridization due to the LS coupling [Eq. (\ref{eq:H_LS})] is essential for the odd-frequency Cooper pairs because they disappear in the absence of the LS coupling.
The representative components of even-parity $A_{1g}$ odd-frequency pair amplitudes at $k_z=0$ are shown in Fig. \ref{fig:Fodd_PE}.  
Because of the time-reversal symmetry, the spin-singlet (spin-triplet) even-parity orbital-singlet (orbital-triplet) odd-frequency pair amplitude $\mathcal{F}^{-+--}_{ll'}$ ($\bm{\mathcal{F}}_{ll'}^{+++-}$) is pure imaginary (real). 
The pair amplitude is normalized by a maximum value of the BCS pair amplitude $\mathcal{F}^{\rm BCS}_{\rm max}(\omega_m)$ which is defined as
\begin{equation}
\mathcal{F}^{\rm BCS}_{\rm max}(\omega_m)\equiv
\underset{\bm{k},l,l'}{\mathrm{max}}
\left|\mathcal{F}_{ll'}^{-+++}(\bm{k},i\omega_m)\right| ,
\end{equation} 
where $\mathcal{F}_{ll'}^{-+++}$ is given by Eq. (\ref{eq:F_-+++}) in Appendix \ref{sec:even-w}. 
As shown in Fig. \ref{fig:Fodd_PE}(a), the spin-singlet even-parity orbital-singlet odd-frequency pair amplitude $\mathcal{F}^{-+--}_{ll'}$ is not generated in our model although it is symmetrically allowed. 
On the other hand, some components of the spin-triplet even-parity orbital-triplet odd-frequency pair amplitudes $\bm{\mathcal{F}}_{ll'}^{+++-}$, that have $d$- or $s+d$-wave like $\bm{k}$-dependence, are generated although their value is small compared to the BCS pair amplitude [Figs. \ref{fig:Fodd_PE}(b) and \ref{fig:Fodd_PE}(c)]. 
These results can be understood on the basis of the group theoretical classification in Sec.\ref{sec2-3}. 
Table \ref{tab:classification_oddw} shows that the $\bm{k}$-dependence of $\mathcal{F}_{ll'}^{-+--}$ is proportional to $k_{\mu}k_{\mu'}$ ($\mu=x,y,z$ and $\mu\neq\mu'$) near the $\Gamma$ point. 
This means that the generation of $\mathcal{F}_{ll'}^{-+--}$ requires the existence of terms proportional to $k_{\mu}k_{\mu'}$ in the model Hamiltonian. 
Indeed, the even-parity interorbital hybridizations proportional to $\sin k_{\mu} \sin k_{\mu'}$ are allowed by symmetry, and they may generate odd-frequency pair amplitudes $\mathcal{F}_{ll'}^{-+--}$. 
However, we omitted such hybridizations since their amplitude is tiny in bulk STO \cite{hirayama2012ab,PhysRevB.86.125121,PhysRevB.87.161102,PhysRevB.100.094504}.  
Then, $\mathcal{F}_{ll'}^{-+--}$ is not induced in our model. 
On the other hand, Table \ref{tab:classification_oddw} also shows that $\bm{k}$-dependence of some components in $\bm{\mathcal{F}}_{ll'}^{+++-}$ are given by a linear combination of $k_{\mu}^2$ (e.g., $\mathcal{F}_{12,z}^{+++-}\sim k_x^2-k_y^2$).  Thus, the generation of these components is expected in bulk STO. 
We found finite $\mathcal{F}_{12,z}^{+++-}$, $\mathcal{F}_{13,y}^{+++-}$, and $\mathcal{F}_{23,x}^{+++-}$ in the PE phase as expected. 
Intraorbital components $\bm{\mathcal{F}}_{ll}^{+++-}$ vanish similarly to $\mathcal{F}_{ll}^{-+--}$. 

\subsection{\label{sec4-2} Ferroelectric phase}
\begin{figure}[tbp]
\centering
   \includegraphics[width=86mm,clip]{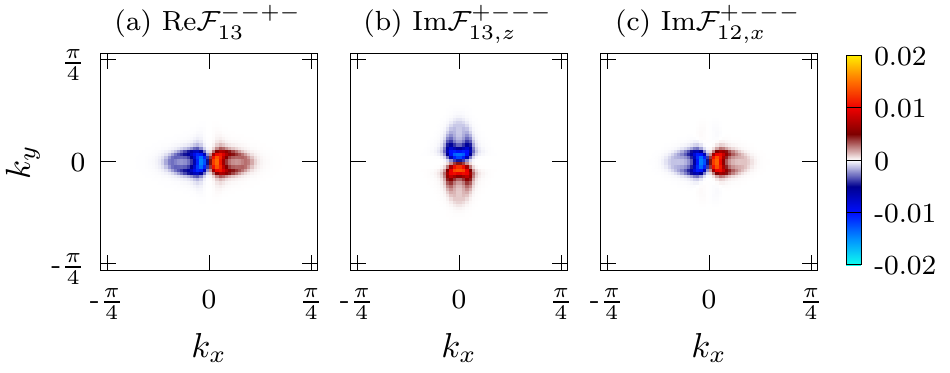}
   \caption{$\bm{k}$-dependence of the odd-parity odd-frequency pair amplitudes (a) $\mathrm{Re}\mathcal{F}_{13}^{--+-}(k)$, (b) $\mathrm{Im}\mathcal{F}_{13,z}^{+---}(k)$, and (c) $\mathrm{Im}\mathcal{F}_{12,x}^{+---}(k)$ at $k_z=0$ in the FE phase. The Matsubara frequency $\omega_m$ is set to be 1 meV, and the values of the pair amplitudes are normalized by $\mathcal{F}^{\rm BCS}_{\rm max}(\omega_m)=0.251432$ meV$^{-1}$. 
 \label{fig:Fodd_A2u} }
\end{figure}

\begin{figure}[tbp]
\centering
   \includegraphics[width=86mm,clip]{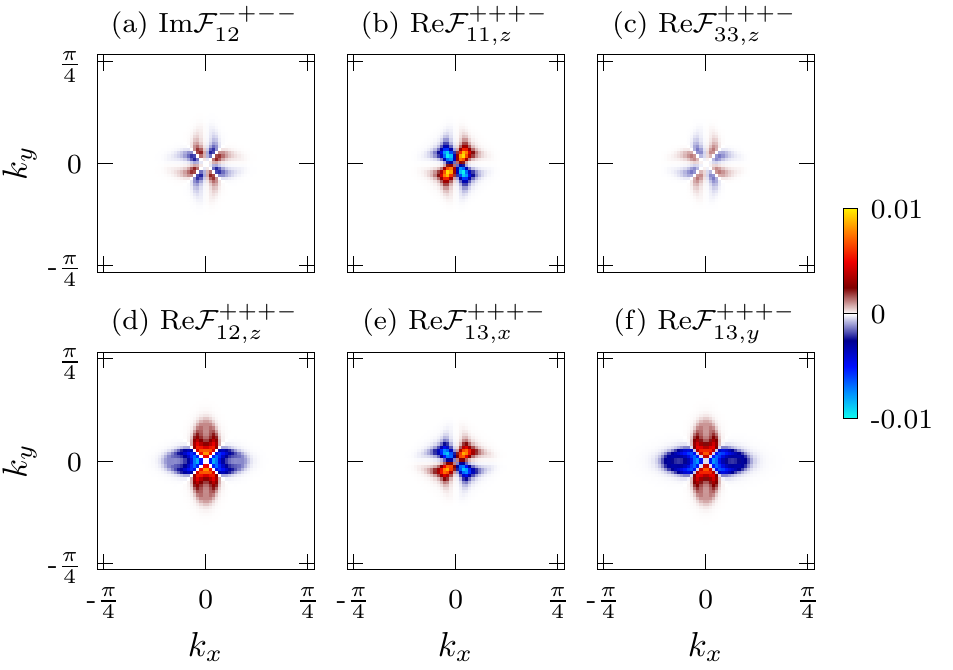}
   \caption{$\bm{k}$-dependence of the even-parity odd-frequency pair amplitudes (a) $\mathrm{Re}\mathcal{F}_{12}^{-+--}(k)$, (b) $\mathrm{Im}\mathcal{F}_{11,z}^{+++-}(k)$, (c) $\mathrm{Im}\mathcal{F}_{33,z}^{+++-}(k)$, (d) $\mathrm{Im}\mathcal{F}_{12,z}^{+++-}(k)$, (e) $\mathrm{Im}\mathcal{F}_{13,x}^{+++-}(k)$, (f) $\mathrm{Im}\mathcal{F}_{13,y}^{+++-}(k)$ at $k_z=0$ in the FE phase. The Matsubara frequency $\omega_m$ is set to be 1 meV, and the values of the pair amplitudes are normalized by $\mathcal{F}^{\rm BCS}_{\rm max}(\omega_m)=0.251432$ meV$^{-1}$. 
 \label{fig:Fodd_A1g} }
\end{figure}

Next, we investigate the odd-frequency pairing in the FE phase with $C_{4v}$ point group symmetry. 
In the FE phase, the odd-parity odd-frequency pair amplitudes, that belong to $A_{2u}$ IR, can also be induced owing to the breakdown of the spacial inversion symmetry. 
Figure \ref{fig:Fodd_A2u} shows the $\bm{k}$-dependence of the representative components of odd-parity $A_{2u}$ odd-frequency pair amplitudes. 
Owing to the time-reversal symmetry, the spin-singlet (spin-triplet) odd-parity orbital-triplet (orbital-singlet) odd-frequency pair amplitude $\mathcal{F}^{--+-}_{ll'}$ ($\bm{\mathcal{F}}_{ll'}^{+---}$) is real (pure imaginary). 
They have $p$-wave like $\bm{k}$-dependence in consistent with the group theoretical classification in Table \ref{tab:classification_oddw}. 
These {\it ferroelectricity-induced} odd-frequency pair amplitudes originate from the odd-parity interorbital hybridization $\zeta_{1,2}(\bm{k})\propto\sin k_{x,y}$ which appear in the FE phase [Eq. (\ref{eq:H_pol})], and hence they should have $p_{x,y}$-wave like $\bm{k}$-dependence (e.g., $p_{z}$-wave like pair amplitude $\mathcal{F}_{ll}^{--+-}\sim k_z$ is not driven by the FE order along the [001] axis, although it is allowed by symmetry). 
We confirmed that $\mathcal{F}_{13}^{--+-}$, $\mathcal{F}_{23}^{--+-}$, $\mathcal{F}_{13,z}^{+---}$, $\mathcal{F}_{23,z}^{+---}$, $\mathcal{F}_{12,x}^{+---}$, and $\mathcal{F}_{12,y}^{+---}$ are induced by the FE transition in bulk STO [Fig. \ref{fig:Fodd_A2u}] in accordance with the group theoretical classification as well as above discussions. 
In addition, the odd-parity interorbital hybridization $\zeta_{l}(\bm{k})$ also generates some components of the even-parity $A_{1g}$ odd-frequency pair amplitudes, whose $\bm{k}$-dependences are proportional to $k_x^{a}k_y^{b}$ ($a,b=0,1,2,\cdots$). 
For instance, $\mathcal{F}_{12}^{-+--}\sim k_xk_y(k_x^2-k_y^2)$ takes a finite value in the FE phase [Fig. \ref{fig:Fodd_A1g}(a)] although it is zero in the PE phase [Fig. \ref{fig:Fodd_PE}(a)]. 
We found that $\mathcal{F}_{12}^{-+--}$, $\mathcal{F}_{11,z}^{+++-}$, $\mathcal{F}_{22,z}^{+++-}$, $\mathcal{F}_{33,z}^{+++-}$, $\mathcal{F}_{12,z}^{+++-}$, $\mathcal{F}_{13,x}^{+++-}$, $\mathcal{F}_{13,y}^{+++-}$, $\mathcal{F}_{23,x}^{+++-}$, and $\mathcal{F}_{23,y}^{+++-}$ are finite in the FE phase [Fig. \ref{fig:Fodd_A1g}]. 

\section{\label{sec5} Experimental signaures of odd-parity orbital hybridization}
In this section, we demonstrate possible experimental signatures of the ferroelectricity-induced odd-parity orbital hybridization $\zeta_{l}(\bm{k})$ that leads to the emergence and enhancement of the odd-frequency pair amplitudes in bulk STO. 

\subsection{\label{sec5-1} Spectral function}
\begin{figure}[tbp]
\centering
   \includegraphics[width=90mm,clip]{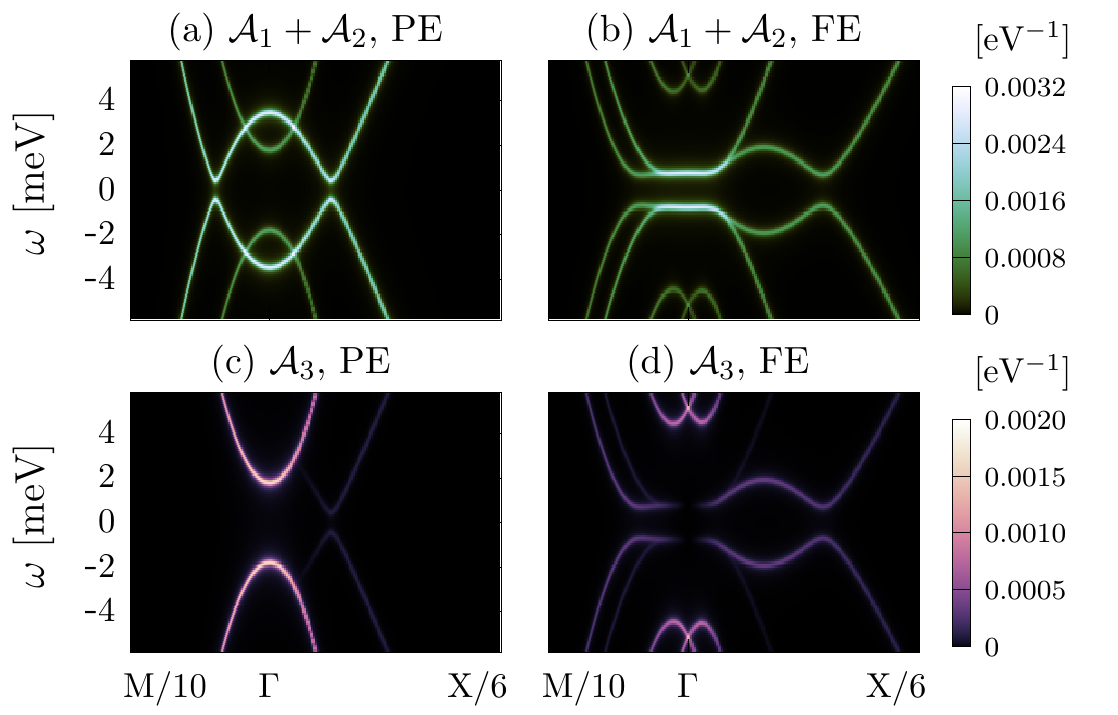}
   \caption{
Orbital-resolved spectral functions (a) $\mathcal{A}_{1}(\bm{k}, \omega)+\mathcal{A}_{2}(\bm{k}, \omega)$ in the PE phase, (b) $\mathcal{A}_{1}(\bm{k}, \omega)+\mathcal{A}_{2}(\bm{k}, \omega)$ in the FE phase, (c) $\mathcal{A}_{3}(\bm{k}, \omega)$ in the PE phase, and (d) $\mathcal{A}_{3}(\bm{k}, \omega)$ in the FE phase. $\Gamma$, $\mathrm{X}$, and $\mathrm{M}$ refer to $\bm{k}=(0,0,0)$, $\bm{k}=(\pi,0,0)$, and $\bm{k}=(\pi,\pi,0)$, respectively. 
 \label{fig:Akw} }
\end{figure}

First, we study the spectral function which can be directly measured by the angle-resolved photoemission spectroscopy (ARPES). 
Using the Matsubara representation of the normal Green's function $\hat{\mathcal{G}}^{0}(k)$ in Eq. (\ref{eq:Matsubara_G}), the spectral function is obtained as 
\begin{equation}
\hat{A}(\bm{k},\omega) = -\frac{1}{\pi} \mathrm{Im}
\hat{\mathcal{G}}^{0}(\bm{k}, \omega+i0^{+}) .
\label{eq:Akw}
\end{equation}
Here, we define the orbital-resolved spectral function as
\begin{equation}
\mathcal{A}_{l}(\bm{k}, \omega) = \sum_{s=\uparrow,\downarrow} A_{ls,ls}(\bm{k},\omega) , 
\end{equation}
whose $(\bm{k}, \omega)$-dependences are shown in Fig. \ref{fig:Akw}. 
The electronic structure near the Fermi level is mainly constructed from $d_{yz}$ and $d_{xz}$ orbitals in the PE phase [Figs. \ref{fig:Akw}(a) and \ref{fig:Akw}(c)]. 
Then, $d_{yz}$ and $d_{xz}$ orbitals are responsible for the superconductivity, and the orbital-resolved spectral function $\mathcal{A}_{1}(\bm{k}, \omega)+\mathcal{A}_{2}(\bm{k}, \omega)$ clearly shows a single superconducting gap at zero energy. 
In the FE phase, the orbital-resolved spectral functions exhibit the Rashba spin-orbit splitting due to the spacial inversion symmetry breaking [Figs. \ref{fig:Akw}(b) and \ref{fig:Akw}(d)]. 
Since the odd-parity hybridization $\zeta_{l}(\bm{k})$ induces the mixing between $d_{yz,xz}$ and $d_{xy}$ orbitals, the weight of $d_{xy}$ orbital at the Fermi level is enhanced by the FE transition [compare Fig. \ref{fig:Akw}(d) with Fig. \ref{fig:Akw}(c)]. 
Thus, the contribution of $d_{xy}$ orbital to the superconductivity is increased in the FE phase.

\begin{figure}[tbp]
\centering
   \includegraphics[width=75mm,clip]{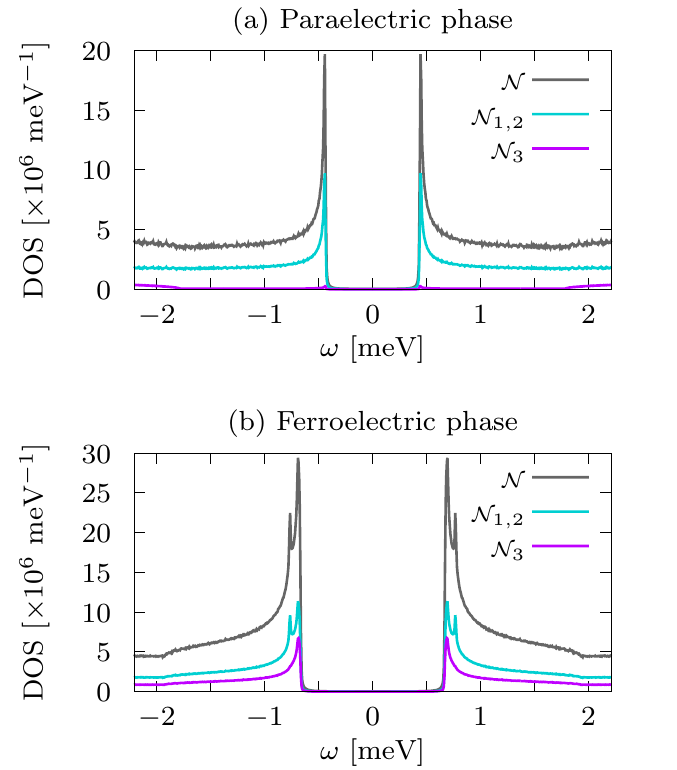}
   \caption{The total DOS $\mathcal{N}(\omega)$ and orbital-resolved DOS $\mathcal{N}_{l}(\omega)$ in the (a) PE phase and (b) FE phase. 
 \label{fig:DOS} }
\end{figure}

\subsection{\label{sec5-2} Density of states}
To see the detail of the superconducting gap structure, we here investigate the DOS which can be measured by tunneling spectroscopy. 
The DOS is obtained by taking the trace of the spectral function as  
\begin{equation}
\mathcal{N}(\omega)=\sum_{\bm{k}} \mathrm{Tr} \hat{A}(\bm{k},\omega)
\equiv\sum_{l} \mathcal{N}_{l}(\omega) ,
\label{eq:DOS_orbital}
\end{equation}
where we defined the orbital-resolved DOS $\mathcal{N}_{l}(\omega)=\sum_{\bm{k}}\mathcal{A}_{l}(\bm{k},\omega)$. 
Figure \ref{fig:DOS} shows the total and orbital-resolved DOS in bulk STO. 
Because of the 4-fold rotational symmetry of the system, the orbital-resolved DOS for $d_{yz}$ orbital $\mathcal{N}_{1}(\omega)$ and that for $d_{xz}$ orbital $\mathcal{N}_{2}(\omega)$ are equivalent. 

In the PE phase, the DOS exhibits the BCS-like form with a single set of coherence peaks [Fig. \ref{fig:DOS}(a)]. 
The coherence peaks originate from $\mathcal{N}_{1,2}(\omega)$ since the electronic states near the Fermi level are mainly constructed from $d_{yz,xz}$ orbitals [see Figs. \ref{fig:Akw}(a) and \ref{fig:Akw}(c)]. 
In the FE phase, the orbital-resolved DOS for $d_{xy}$ orbital $\mathcal{N}_{3}(\omega)$ becomes comparable to $\mathcal{N}_{1,2}(\omega)$ owing to the odd-parity hybridization in the FE phase [Fig. \ref{fig:DOS}(b)]. 
The total DOS shows two sets of the coherence peaks in the FE phase. 
The lower energy coherence peaks originate from both $\mathcal{N}_{1,2}(\omega)$ and $\mathcal{N}_{3}(\omega)$, and hence they are generated by all of $t_{2g}$ orbitals. 
On the other hand, the higher energy coherence peaks originate only from $\mathcal{N}_{1,2}(\omega)$. 
We also note that the DOS shows the ferroelectricity-induced enhancement of superconducting gap which originates from the Lifshitz transition \cite{PhysRevB.100.094504}. 

\subsection{\label{sec5-3} Gap functions in band basis}
\begin{figure}[tbp]
\centering
   \includegraphics[width=65mm,clip]{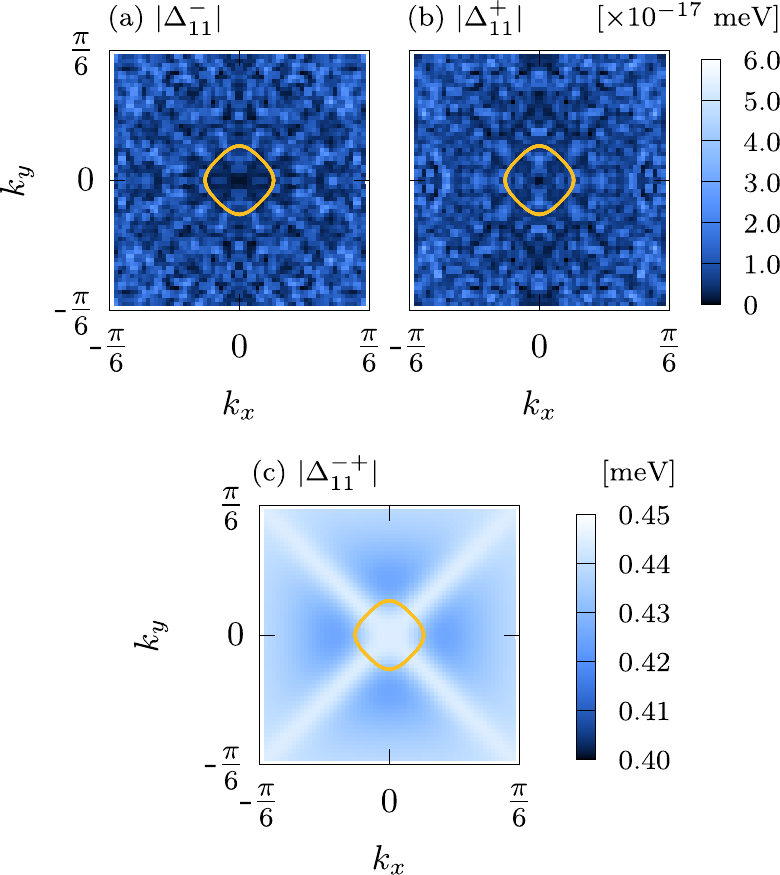}
   \caption{$\bm{k}$-dependence of the superconducting gap functions (a) $|\Delta_{11}^{-}(\bm{k})|$, (b) $|\Delta_{11}^{+}(\bm{k})|$, and (c) $|\Delta_{11}^{-+}(\bm{k})|$ at $k_z=0$ in the PE phase. The yellow lines show the Fermi surface. 
 \label{fig:Gap_PE} }
\end{figure}

\begin{figure}[tbp]
\centering
   \includegraphics[width=65mm,clip]{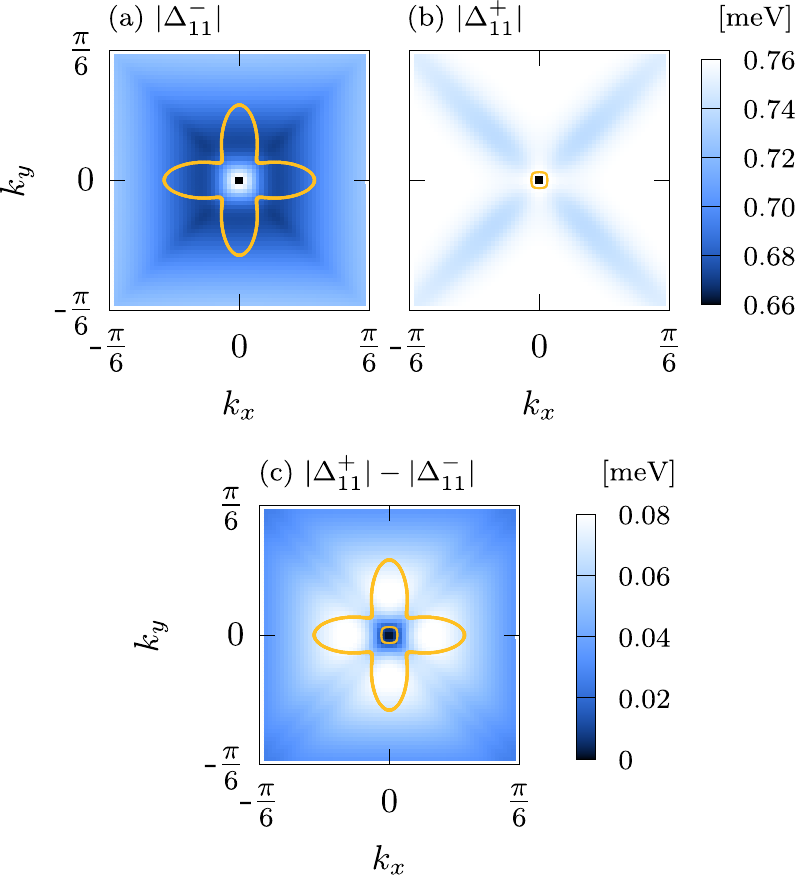}
   \caption{$\bm{k}$-dependence of the the superconducting gap functions (a) $|\Delta_{11}^{-}(\bm{k})|$, (b) $|\Delta_{11}^{+}(\bm{k})|$, and (c) their difference $|\Delta_{11}^{+}(\bm{k})|-|\Delta_{11}^{-}(\bm{k})|$ at $k_z=0$ in the FE phase. The yellow lines show the corresponding Fermi surfaces. 
 \label{fig:Gap_FE} }
\end{figure}

To clarify the ferroelectricity-induced splitting of coherence peaks in the DOS, we here investigate the superconducting gap functions in the band basis. 
We carry out the band diagonalization for the BdG Hamiltonian as follows: 
\begin{align}
\hat{U}(\bm{k}) \hat{\mathcal{H}}(\bm{k}) \hat{U}^{\dag}(\bm{k}) = 
\begin{pmatrix}
 \hat{\mathcal{E}}_{0}(\bm{k}) & \hat{\Delta}_{\rm band}(\bm{k}) \\
 \hat{\Delta}_{\rm band}^{\dag}(\bm{k}) & -\hat{\mathcal{E}}_{0}(-\bm{k})
\end{pmatrix} , 
\end{align}
where the unitary matrix $\hat{U}(\bm{k})$ is defined as
\begin{align}
\hat{U}(\bm{k}) = 
\begin{pmatrix}
 \hat{U}_{0}(\bm{k}) & \hat{0} \\
 \hat{0} & \hat{U}_{0}^{*}(-\bm{k})
\end{pmatrix} .
\end{align}
The unitary matrix $\hat{U}_{0}(\bm{k})$ diagonalizes the normal state part of the BdG Hamiltonian $\hat{\mathcal{H}}_{0}(\bm{k})$, and then the $6\times 6$ diagonal submatrix $\hat{\mathcal{E}}_{0}(\bm{k})$ is obtained as
\begin{align}
\hat{\mathcal{E}}_{0}(\bm{k}) &= 
\begin{pmatrix}
 \hat{\mathcal{E}}_{1}(\bm{k}) & \hat{0} & \hat{0} \\
 \hat{0} & \hat{\mathcal{E}}_{2}(\bm{k}) & \hat{0} \\
 \hat{0} & \hat{0} & \hat{\mathcal{E}}_{3}(\bm{k}) 
\end{pmatrix} , \\
\hat{\mathcal{E}}_{j}(\bm{k}) &= 
\begin{pmatrix}
 \mathcal{E}_{j}^{-}(\bm{k}) & 0 \\
 0 & \mathcal{E}_{j}^{+}(\bm{k})
\end{pmatrix} .
\end{align} 
Here, $\mathcal{E}_{j}^{\pm}(\bm{k})$ denotes the energy spectrum in the normal state [$\mathcal{E}_{j}^{\pm}(\bm{k})\leq\mathcal{E}_{j'}^{\pm}(\bm{k})$ ($j\leq j'$) and $\mathcal{E}_{j}^{-}(\bm{k})\leq\mathcal{E}_{j}^{+}(\bm{k})$]. 
In the PE phase, the energy spectrum is 2-fold degenerate (i.e., $\mathcal{E}_{j}^{-}(\bm{k})=\mathcal{E}_{j}^{+}(\bm{k})$) owing to the spacial inversion and time-reversal symmetries. 
The pairing part of the BdG Hamiltonian in the band basis $\hat{\Delta}_{\rm band}(\bm{k})$ can be described as
\begin{align}
\hat{\Delta}_{\rm band}(\bm{k}) &= 
\begin{pmatrix}
 \hat{\Delta}_{11}(\bm{k}) & \hat{\Delta}_{12}(\bm{k}) & \hat{\Delta}_{13}(\bm{k}) \\
\hat{\Delta}_{21}(\bm{k}) & \hat{\Delta}_{22}(\bm{k}) & \hat{\Delta}_{23}(\bm{k}) \\
\hat{\Delta}_{31}(\bm{k}) & \hat{\Delta}_{32}(\bm{k}) & \hat{\Delta}_{33}(\bm{k}) 
\end{pmatrix} , \\
\hat{\Delta}_{ij}(\bm{k}) &= 
\begin{pmatrix}
 \Delta_{ij}^{-}(\bm{k}) & \Delta_{ij}^{-+}(\bm{k}) \\
 \Delta_{ij}^{+-}(\bm{k}) & \Delta_{ij}^{+}(\bm{k})
\end{pmatrix} .
\end{align} 
In our calculations, the superconductivity is dominated by the two lowest energy bands that construct the Fermi surfaces. 
Then, the effective two-band Hamiltonian for the two lowest energy bands $\mathcal{E}_{1}^{\pm}(\bm{k})$ can be obtained as
\begin{align}
\hat{\mathcal{H}}_{\rm eff}(\bm{k}) = 
\begin{pmatrix}
 \hat{\mathcal{E}}_{1}(\bm{k}) & \hat{\Delta}_{11}(\bm{k}) \\
 \hat{\Delta}_{11}^{\dag}(\bm{k}) & -\hat{\mathcal{E}}_{1}(-\bm{k})
\end{pmatrix} ,
\end{align} 
by neglecting the higher energy bands that do not give rise to the Fermi surface. 
Figure \ref{fig:Gap_PE} shows the superconducting gap functions and 2-fold degenerate Fermi surfaces in the PE phase. 
It is shown that the pseudospin-singlet superconducting state is stabilized (i.e., $\Delta_{11}^{\pm}(\bm{k})=0$ [Figs. \ref{fig:Gap_PE}(a) and \ref{fig:Gap_PE}(b)] and $\Delta_{11}^{-+}(\bm{k})=-\Delta_{11}^{+-}(\bm{k})$). 
Thus, the superconducting gap in the PE phase is given by $2|\Delta_{11}^{-+}(\bm{k})|$, and its magnitude is consistent with the zero energy gap in the DOS [compare Fig. \ref{fig:DOS}(a) with Fig. \ref{fig:Gap_PE}(c)]. 
On the other hand, Fig. \ref{fig:Gap_FE} shows the superconducting gap functions and Rashba-split Fermi surfaces in the FE phase. 
It is shown that the intraband superconducting state is stabilized as $\Delta_{11}^{\pm}(\bm{k})\neq0$ and $\Delta_{11}^{\mp\pm}(\bm{k})\simeq0$.  Then, the superconducting gaps for each Fermi surfaces are given by $2|\Delta_{11}^{\pm}(\bm{k})|$. 
The difference between $\Delta_{11}^{+}(\bm{k})$ and $\Delta_{11}^{-}(\bm{k})$ leads to a two gap structure shown in Fig. \ref{fig:DOS}(b). 
The two gap structure does not appear in single-orbital Rashba systems (see Appendix \ref{sec:Rashba}), and it is indeed a fingerprint of multiorbital Rashba superconductors. 

Now, we elucidate the origin of  the two coherence peaks in the FE phase. 
The maximum value of $|\Delta_{11}^{+}(\bm{k})|-|\Delta_{11}^{-}(\bm{k})|$ is approximately equivalent to the splitting width of the two superconducting gaps  [compare Fig. \ref{fig:DOS}(b) and Fig. \ref{fig:Gap_FE}(c)]. 
Therefore, the splitting of the coherence peaks in DOS can be attributed to the difference of the superconducting gap between two Rashba-split Fermi surfaces. 
The difference between $|\Delta_{11}^{+}(\bm{k})|$ and $|\Delta_{11}^{-}(\bm{k})|$ originates from their different orbital character. 
The orbital-resolved spectral function $\mathcal{A}_{3}(\bm{k},\omega)$ reveals that $d_{xy}$ orbital does not contribute to the superconducting gap opening at $\Gamma$-point in the FE phase [see Fig. \ref{fig:Akw}(d)], since the odd-parity hybridization vanishes at $k_{x,y}=0$ (i.e., $\zeta_{1}(0,k_y,k_z)=\zeta_{2}(k_x,0,k_z)=0$). 
On the other hand, the Fermi surface for the higher (lower) energy band $\mathcal{E}_{1}^{+}(\bm{k})$ ($\mathcal{E}_{1}^{-}(\bm{k})$) is located near the $\Gamma$-point (away from the $\Gamma$-point). 
Then, $\Delta_{11}^{+}(\bm{k})$ is mainly constructed from $d_{yz,xz}$ orbitals, while $\Delta_{11}^{-}(\bm{k})$ has contribution from $d_{xy}$ orbital comparable to that from $d_{yz,xz}$ orbitals. 
Therefore, $|\Delta_{11}^{+}(\bm{k})|>|\Delta_{11}^{-}(\bm{k})|$ is realized [Figs. \ref{fig:Gap_FE}(a) and \ref{fig:Gap_FE}(b)] because the superconducting order parameter for $d_{yz,xz}$ orbitals is larger than that for $d_{xy}$ orbital (i.e., $\Delta_{1,2} > \Delta_{3}$) in our calculation [see Table \ref{tab:FESC}]. 
Indeed, the higher energy coherence peaks at $\pm\Delta_{11}^{+}(\bm{k})$ for the small Fermi surface are constructed only from $d_{yz,xz}$ orbitals. 
We note that the two gap structure discussed above is essentially different from that due to parity mixing in Cooper pairs. 


\section{\label{sec6} Summary and conclusion}
In summary, we have studied the odd-frequency pair correlations in bulk STO near a FE critical point, based on the group theoretical classification and microscopic numerical calculations. 
First, by considering the fermionic and point group symmetries of a $t_{2g}$ electron system, we have provided a classification table for the odd-frequency pair amplitudes. 
The $\bm{k}$-dependence for both intraorbital and interorbital pair amplitudes has been clarified. 
By combining with the symmetry of terms of the normal state Hamiltonian, the classification table enables us to predict which components of the odd-frequency pair correlations are generated. 
Then, we have calculated the odd-frequency pair amplitudes in a dilute carrier density regime of STO, by using a three-orbital tight-binding model coupled to polar lattice distortion. 
The obtained results are consistent with the group theoretical classification. 

In the PE phase, the spin-triplet even-parity orbital-triplet odd-frequency pair correlations, that belong to $A_{1g}$ IR of $D_{4h}$ point group, are generated owing to the intrinsic LS coupling which leads to local orbital hybridization. 
In the FE phase, additional odd-parity odd-frequency pair correlations, that belong to $A_{2u}$ IR of $D_{4h}$ point group, are induced due to the odd-parity orbital mixing term proportional to the FE order parameter. 
This odd-parity orbital hybridization also leads to the generation of the spin-singlet even-parity orbital-singlet odd-frequency pair correlations. 
We have demonstrated that experimental signatures of the odd-parity orbital hybridization, which is essential for the ferroelectricity-induced odd-frequency pair correlations, can be observed in the spectral functions and DOS. 
In the FE phase, the orbital-resolved spectral function exhibits the Rashba splitting, and the weight of $d_{xy}$ orbital at the Fermi level is increased.  
Accordingly, the DOS in the FE superconducting phase shows a two gap structure, while that in the PE superconducting phase has a BCS-like single gap structure.  
The splitting of the coherence peaks in the FE phase is attributed to the difference in the orbital character between Rashba split Fermi surfaces. 
This is indeed a characteristic property of the multiorbital Rashba superconductors. 

This paper is a first proposal of the multiorbital odd-frequency superconductivity in bulk STO. 
Since we assumed an intraorbital superconducting order parameter, the essential ingredient for the generation of the odd-frequency pair correlations is the orbital mixing in the normal state \cite{BlackSchaffer2013}, which comes from the intrinsic LS coupling and polar inversion symmetry breaking. 
In addition, as far as we know, we have demonstrated a first example of group theoretical classification for the odd-frequency superconducting state in multiorbital systems. 
Our classification method is based on both fermionic symmetry of the anmalous Green's function (Berezinskii rule) and space group symmetry of the system, and may provide a useful tool for searching the odd-frequency pair correlations in multiorbital systems. 

\begin{acknowledgments}
The authors are grateful to A. M. Black-Schaffer and J. Haraldsen for helpful discussions. 
This work was supported by VR 2017-03997, and by KAW 2018-0104, and by JSPS KAKENHI (Grants No. JP15H05884, No. JP18H04225, No. JP18H05227, No. JP18H01178, and No. 20H05159).  
S. K. is supported by a JSPS research fellowship and by JSPS KAKENHI (Gran No. 19J22122). 
\end{acknowledgments}

\appendix
\section{\label{sec:even-w} Even-frequency pair correlations}
In the main text, we have classified the odd-frequency pair amplitude based on the fermionic symmetry and space group symmetry. 
Here, we classify the even-frequency pair amplitude in the same manner. 
By using Eqs. (\ref{eq:singlet}) and (\ref{eq:triplet}), the even-frequency pair amplitudes, that are even under the time inversion $\mathcal{T}$, can be obtained as 
\begin{align}
\mathcal{F}_{ll'}^{-+++}&=\frac{(\psi_{ll'}^{+}+\mathcal{P}\psi_{ll'}^{+})+\mathcal{T}(\psi_{ll'}^{+}+\mathcal{P}\psi_{ll'}^{+})}{4} , \label{eq:F_-+++} \\
\mathcal{F}_{ll'}^{---+}&=\frac{(\psi_{ll'}^{-}-\mathcal{P}\psi_{ll'}^{-})+\mathcal{T}(\psi_{ll'}^{-}-\mathcal{P}\psi_{ll'}^{-})}{4} , \label{eq:F_---+}  \\
\bm{\mathcal{F}}_{ll'}^{++-+}&=\frac{(\bm{d}_{ll'}^{-}+\mathcal{P}\bm{d}_{ll'}^{-})+\mathcal{T}(\bm{d}_{ll'}^{-}+\mathcal{P}\bm{d}_{ll'}^{-})}{4} , \label{eq:F_++-+}  \\
\bm{\mathcal{F}}_{ll'}^{+-++}&=\frac{(\bm{d}_{ll'}^{+}-\mathcal{P}\bm{d}_{ll'}^{+})+\mathcal{T}(\bm{d}_{ll'}^{+}-\mathcal{P}\bm{d}_{ll'}^{+})}{4} . \label{eq:F_+-++} 
\end{align}
The above even-frequency pair amplitudes satisfy the Berezinskii rule ($\mathcal{SPOT}=-1$) shown in Table \ref{tab:classification_SPOT}. 
Then, the basis functions for $\bm{k}$-dependence of the even-frequency pair amplitudes are determined by using Eq. (\ref{eq:gap_Gtrans}). 
The result of the classification is summarized in Table \ref{tab:classification_evenw}.

\begin{table*}[htbp] 
\caption{\label{tab:classification_evenw}
Basis functions for $\bm{k}$-dependence of the even-frequency pair amplitude in a $t_{2g}$ electron system under $D_{4h}$ point group symmetry. 
Basis functions of $A_{1g}$ and $A_{2u}$ IRs are listed, because the superconducting gap function is assumed to belong the trivial IR of $D_{4h}$ or $C_{4v}$. 
The orbital index $l=1,2,3$ refers to $d_{yz}, d_{xz}, d_{xy}$ orbitals, respectively. $\phi_{j+}^{\Gamma}$ and $\eta_{j+}^{\Gamma}$ ($\Gamma=A_{1g}, A_{2u}$ and $j=1,2,\cdots$) are arbitrary constants. 
}
\centering
\begin{ruledtabular}
{\renewcommand \arraystretch{1.4}
 \begin{tabular}{ccc} 
	$\mathcal{F}_{ll'}^{-+++}$  & $\bm{k}$-dependence of $\mathcal{F}_{ll'}^{-+++}(\bm{k},i\omega_m)$ & IR  \\ \hline 
  $\mathcal{F}_{11}^{-+++}$
     & $\phi^{A_{1g}}_{1+}+\phi^{A_{1g}}_{2+}k_x^2+\phi^{A_{1g}}_{3+}k_y^2+\phi^{A_{1g}}_{4+}k_z^2$  &  \multirow{6}{*}{$A_{1g}$} \\  
  $\mathcal{F}_{22}^{-+++}$
     & $\phi^{A_{1g}}_{1+}+\phi^{A_{1g}}_{3+}k_x^2+\phi^{A_{1g}}_{2+}k_y^2+\phi^{A_{1g}}_{4+}k_z^2$  \\
  $\mathcal{F}_{33}^{-+++}$
     & $\phi^{A_{1g}}_{5+}+\phi^{A_{1g}}_{6+}(k_x^2+k_y^2)+\phi^{A_{1g}}_{7+}k_z^2$  \\
  $\mathcal{F}_{12}^{-+++}$
     & $\phi^{A_{1g}}_{8+}k_xk_y$  \\
  $\mathcal{F}_{13}^{-+++}$
     & $\phi^{A_{1g}}_{9+}k_xk_z$  \\
  $\mathcal{F}_{23}^{-+++}$
     & $\phi^{A_{1g}}_{9+}k_yk_z$  \\ \hline\hline
  $\mathcal{F}_{ll'}^{---+}$  & $\bm{k}$-dependence of $\mathcal{F}_{ll'}^{---+}(\bm{k},i\omega_m)$ & IR  \\ \hline 
  $\mathcal{F}_{12}^{---+}$
			& $\phi^{A_{2u}}_{1+}k_xk_yk_z(k_x^2-k_y^2)$ & \multirow{3}{*}{$A_{2u}$} \\ 
  $\mathcal{F}_{13}^{---+}$
        & $\phi^{A_{2u}}_{2+}k_x$ \\ 
  $\mathcal{F}_{23}^{---+}$
        & $\phi^{A_{2u}}_{2+}k_y$ \\ \hline\hline
$\bm{\mathcal{F}}_{ll'}^{++-+}$  & $\bm{k}$-dependence of $\bm{\mathcal{F}}_{ll'}^{++-+}(\bm{k},i\omega_m)$ & IR  \\ \hline 
  $\bm{\mathcal{F}}_{12}^{++-+}$ 
   & $\eta^{A_{1g}}_{1+}(k_xk_z\hat{\bm{x}}+k_yk_z\hat{\bm{y}})+[\eta^{A_{1g}}_{2+}+\eta^{A_{1g}}_{3+}(k_x^2+k_y^2)+\eta^{A_{1g}}_{4+}k_z^2]\hat{\bm{z}}$ & \multirow{3}{*}{$A_{1g}$} \\ 
  $\bm{\mathcal{F}}_{13}^{++-+}$ 
   & $\eta^{A_{1g}}_{5+}k_xk_y\hat{\bm{x}}+(\eta^{A_{1g}}_{6+}+\eta^{A_{1g}}_{7+}k_x^2+\eta^{A_{1g}}_{8+}k_y^2+\eta^{A_{1g}}_{9+}k_z^2)\hat{\bm{y}}+\eta^{A_{1g}}_{10+}k_yk_z\hat{\bm{z}}$  \\
  $\bm{\mathcal{F}}_{23}^{++-+}$ 
   & $-(\eta^{A_{1g}}_{6+}+\eta^{A_{1g}}_{8+}k_x^2+\eta^{A_{1g}}_{7+}k_y^2+\eta^{A_{1g}}_{9+}k_z^2)\hat{\bm{x}}-\eta^{A_{1g}}_{5+}k_xk_y\hat{\bm{y}}-\eta^{A_{1g}}_{10+}k_xk_z\hat{\bm{z}}$ \\ \hline\hline
$\bm{\mathcal{F}}_{ll'}^{+-++}$  & $\bm{k}$-dependence of $\bm{\mathcal{F}}_{ll'}^{+-++}(\bm{k},i\omega_m)$ & IR  \\ \hline 
  $\bm{\mathcal{F}}_{11}^{+-++}$
      & $\eta^{A_{2u}}_{1+}k_y\hat{\bm{x}}+\eta^{A_{2u}}_{2+}k_x\hat{\bm{y}}+\eta^{A_{2u}}_{3+}k_xk_yk_z\hat{\bm{z}}$ & \multirow{6}{*}{$A_{2u}$} \\ 
  $\bm{\mathcal{F}}_{22}^{+-++}$
      & $-\eta^{A_{2u}}_{2+}k_y\hat{\bm{x}}-\eta^{A_{2u}}_{1+}k_x\hat{\bm{y}}-\eta^{A_{2u}}_{3+}k_xk_yk_z\hat{\bm{z}}$ \\
  $\bm{\mathcal{F}}_{33}^{+-++}$
      & $\eta^{A_{2u}}_{4+}(k_y\hat{\bm{x}}-k_x\hat{\bm{y}})+\eta^{A_{2u}}_{5+}k_xk_yk_z(k_x^2-k_y^2)\hat{\bm{z}}$ \\
  $\bm{\mathcal{F}}_{12}^{+-++}$
      & $\eta^{A_{2u}}_{6+}(k_x\hat{\bm{x}}-k_y\hat{\bm{y}})+\eta^{A_{2u}}_{7+}k_z(k_x^2-k_y^2)\hat{\bm{z}}$  \\
  $\bm{\mathcal{F}}_{13}^{+-++}$
      & $\eta^{A_{2u}}_{8+}k_xk_yk_z\hat{\bm{x}}+\eta^{A_{2u}}_{9+}k_z\hat{\bm{y}}+\eta^{A_{2u}}_{10+}k_y\hat{\bm{z}}$  \\
  $\bm{\mathcal{F}}_{23}^{+-++}$
      & $-\eta^{A_{2u}}_{9+}k_z\hat{\bm{x}}-\eta^{A_{2u}}_{8+}k_xk_yk_z\hat{\bm{y}}-\eta^{A_{2u}}_{10+}k_x\hat{\bm{z}}$ 
\end{tabular}
}
\end{ruledtabular}
\end{table*} 

\section{\label{sec:Rashba} Single-orbital Rashba superconductor}
In this Appendix, we discuss the superconducting gap structure of single-orbital Rashba superconductors to illuminate multiorbital effect in STO. 
We consider a Rashba electron system in the cubic crystal lattice structure for comparison with the model of bulk STO. 
The model Hamiltonian is given by
\begin{eqnarray}
\mathcal{H}=&&\sum_{\bm{k},s}\xi(\bm{k})c_{\bm{k},s}^{\dag}c_{\bm{k},s}+\sum_{\bm{k},s,s'}\alpha\bm{g}(\bm{k})\cdot\bm{\sigma}_{ss'}c_{\bm{k},s}^{\dag}c_{\bm{k},s'} \nonumber\\
&& +\frac{1}{2} \sum_{\bm{k},s,s'}\Delta_{ss'}(\bm{k})c_{\bm{k},s}^{\dag}c_{-\bm{k},s'}^{\dag}+\mathrm{H.c.},
\end{eqnarray}
where $c_{\bm{k},s}$ is the annihilation operator for an electron with momentum $\bm{k}$ and spin $s=\uparrow,\downarrow$. 
The single-electron kinetic energy is described as $\xi(\bm{k})=-2t(\cos k_x+\cos k_y+\cos k_z)-\mu$. 
The Rashba spin-orbit coupling takes the form $\bm{g}(\bm{k})=\sin k_y\hat{\bm{x}}-\sin k_x\hat{\bm{y}}$. 
We assume the superconducting gap function as
\begin{align}
\Delta_{ss'}(\bm{k})= \left[\left(\Delta^{s}\sigma^{0} + \Delta^{p} \bm{g}(\bm{k})\cdot\bm{\sigma} \right) i\sigma^{y} \right]_{ss'} , 
\end{align}
where $\Delta^{s}$ is the spin-singlet $s$-wave component of the gap function, and $\Delta^{p}$ is the spin-triplet $p$-wave component. 
By diagonalizing the BdG Hamiltonian, we obtain the energy spectrum as follows: 
\begin{align}
E(\bm{k}) &= \pm \sqrt{ \mathcal{E}_{\pm}(\bm{k})^2 + |\Delta_{\pm}(\bm{k})|^2 }, 
\end{align}
where $\mathcal{E}_{\pm}(\bm{k})=\xi(\bm{k})\pm\alpha|\bm{g}(\bm{k})|$ and the superconducting gap functions in the band basis are given by
\begin{align}
\Delta_{-}(\bm{k})&=\frac{-g_{x}(\bm{k})+ig_y(\bm{k})}{|\bm{g}(\bm{k})|}\left(\Delta^s-\Delta^p|\bm{g}(\bm{k})|\right) , \\
\Delta_{+}(\bm{k})&=\frac{g_{x}(\bm{k})+ig_y(\bm{k})}{|\bm{g}(\bm{k})|}\left(\Delta^s+\Delta^p|\bm{g}(\bm{k})|\right) .
\end{align}
In the absence of the parity-mixing-induced $p$-wave pairing interaction ($\Delta^{p}=0$), the superconducting gaps for each Rashba split bands are equal ($|\Delta_{-}(\bm{k})|=|\Delta_{+}(\bm{k})|=|\Delta^s|$). 
Therefore, the multiple superconducting gap structure such as in Fig. \ref{fig:DOS}(b) is not realized in conventional $s$-wave Rashba superconductors. 
This is in sharp contrast to the effectively single-band Rashba superconductor discussed in Sec. \ref{sec5-3}, where the two gap structure is induced by the orbital degree of freedom.


\nocite{*}

\providecommand{\noopsort}[1]{}\providecommand{\singleletter}[1]{#1}%

\end{document}